\documentclass[10pt,journal,compsoc]{IEEEtran}
\usepackage{amsmath}
\usepackage{amssymb}
\usepackage{amsfonts}
\usepackage[dvips]{graphicx}

\usepackage{algorithmic}
\usepackage{algorithm}
\usepackage{subfigure}
\usepackage{multicol}
\usepackage{comment}
\newcommand{\argmin}{\mathop{\rm argmin}\limits}
\newcommand{\argmax}{\mathop{\rm argmax}\limits}

\usepackage{eqparbox}
\ifCLASSOPTIONcompsoc
  \usepackage[nocompress]{cite}
\else
  \usepackage{cite}
\fi
\ifCLASSINFOpdf
\else
\fi
\begin{document}
\title{\huge{Optimization of Indexing Based on k-Nearest Neighbor Graph for Proximity Search in High-dimensional Data}}

\author{Masajiro~Iwasaki and
        Daisuke~Miyazaki% <-this % stops a space
\thanks{The authors are with Yahoo Japan Corporation, Tokyo, JAPAN.}}

\IEEEtitleabstractindextext{%
\begin{abstract}
Searching for high-dimensional vector data with high accuracy is an inevitable search technology for various types of data. Graph-based indexes are known to reduce the query time for high-dimensional data. To further improve the query time by using graphs, we focused on the indegrees and outdegrees of graphs. While a sufficient number of incoming edges (indegrees) are indispensable for increasing search accuracy, an excessive number of outgoing edges (outdegrees) should be suppressed so as to not increase the query time. Therefore, we propose three degree-adjustment methods: static degree adjustment of not only outdegrees but also indegrees, dynamic degree adjustment with which outdegrees are determined by the search accuracy users require, and path adjustment to remove edges that have alternative search paths to reduce outdegrees. We also show how to obtain optimal degree-adjustment parameters and that our methods outperformed previous methods for image and textual data.
\end{abstract}

%\begin{IEEEkeywords}
%Computer Society, IEEE, IEEEtran, journal, \LaTeX, paper, template.
%\end{IEEEkeywords}
}

\maketitle

%\IEEEdisplaynontitleabstractindextext

% \fi
%\IEEEpeerreviewmaketitle

\section{Introduction and Related Work}
To search for various types of data, e.g., document, image, audio, or a mixture of them, simple vector objects, which are extracted from original data, are used for searching. Since such vector objects should adequately represent the original data, they tend to be high-dimensional. However, it is difficult to accelerate proximity searches for such high-dimensional objects while maintaining a high search accuracy. There are two types of methods for approximate proximity searches. One type includes hash \cite{gionis1999similarity}\cite{datar2004locality}\cite{weiss2009spectral}, quantization \cite{gong2011iterative}\cite{jegou2011product}\cite{DBLP:journals/corr/JohnsonDJ17}, and permutation-based \cite{esuli2009pp}\cite{tellez2013succinct} methods, which do not require any objects to compute distances to a query object during the search process. The other type includes tree- and graph-based methods, which do require objects. Thus, the former requires less memory than the latter. However, its accuracy tends to be worse. Most applications require better search accuracy. Solid-state drives (SSDs) can be used instead of main memory as storage for objects since high-speed SSDs have been becoming widespread. We therefore focus on the latter to obtain a high search accuracy rather than save on memory usage.

Proximity searches using objects are broadly classified into tree-based and graph-based methods. In tree-based methods, an entire space is hierarchically and recursively divided into subspaces. Various tree-based methods have been proposed, including the kd-tree \cite{bentley1975multidimensional} and vp-tree \cite{yianilos1993data}. While these methods can provide exact search results, tree-based approximate search methods have also been studied to shorten the query time. ANN \cite{arya1998optimal} is a method that applies an approximate search to a kd-tree, and FLANN \cite{muja2014scalable} is an open-source library for approximate proximity searches that provides randomized kd-trees wherein multiple kd-trees are searched in parallel \cite{silpa2008optimised}\cite{muja2014scalable} and k-means trees are constructed by hierarchical k-means partitioning \cite{nister2006scalable}\cite{muja2014scalable}.

Graph-based methods use a neighborhood graph as a search index. Arya et al. \cite{Arya:1993:ANN:313559.313768} proposed a method of using randomized neighbor graphs as a search index. SASH \cite{houle2005fast}, although it has a tree shape, is actually a graph-based method due to its node connections. Sebastian et al. \cite{sebastian2002metric} used a k-nearest neighbor graph (KNNG) as a search index, where each node in the KNNG has directed edges to the k-nearest neighboring nodes. Although a KNNG is a simple graph, it can reduce the query time and provide a high search accuracy. Wang et al. \cite{Wang:2012:QIN:2393347.2393378} improved the query time by using seed nodes (starting nodes for exploring a graph) obtained with a tree-based index depending on the query from an object set. Hajebi et al. \cite{hajebi2011fast} showed that searches using KNNGs outperform LSH and kd-trees for image descriptors. DRNG \cite{aoyama2011fast} reduces the degrees of a KNNG to improve query time. However, as the number of objects grows, the brute force construction cost of a KNNG exponentially increases because the distances between all pairs of objects in a graph need to be computed. To solve this, SW-graph \cite{malkov2014approximate} and ANNG \cite{iwasaki2010E} used approximate neighborhood graphs, where a graph is incrementally constructed from neighboring nodes searched by using a partially constructed graph. The KGraph\footnote{https://github.com/aaalgo/kgraph} library also uses an approximate KNNG \cite{dong2011efficient}. HNSW \cite{DBLP:journals/corr/MalkovY16} has several layers of approximate neighborhood graphs with long edges reaching out to further nodes. These graphs can drastically reduce construction costs while maintaining a short query time. PANNG \cite{iwasaki2016pruned}, which prunes the edges of each node in an ANNG to shorten the query time, outperformed a quantization-based method \cite{jegou2011product}.

When it comes to graph-based indexing, the indexing is difficult to mathematically analyze. Therefore, to precisely analyze it, we focused on indegrees and outdegrees, which are the numbers of the incoming and outgoing edges of each node, respectively, in a graph. During the search process, to reach nodes neighboring a query object, the incoming edges of these nodes are indispensable. Thus, it is assumed that incoming edges play a more important role for a search than outgoing edges in increasing search accuracy. However, excessive edges increase the number of evaluated nodes, increasing the query time. Therefore, we individually adjust not only the outdegrees but also the indegrees unlike previous methods in order to construct an optimal graph. We propose three different types of degree-adjustment methods. First, we developed a static degree-adjustment method for deriving an adjusted graph from the edges and reversed edges of a KNNG to roughly adjust the indegrees and outdegrees. However, this causes some nodes to have a high outdegree, which increases the query time. Thus, we also developed a degree-adjustment method with constraints for more precisely adjusting the indegrees and outdegrees. Although these two methods can construct statically adjusted graphs, the graphs should be adjusted to a level of search accuracy that users require to further improve search performance. Thus, we propose a dynamic degree-adjustment method for dynamically determining the optimal outdegree from the required accuracy at the beginning of a search considering that individual users require different levels of accuracy for each search. These three types take into account only the indegree and outdegree for each node. Paths to a query that are explored during the search process should be optimized as well. Therefore, we also propose a path-adjustment method for unnecessary shortcut edges that have alternative paths to drastically reduce degrees. However, the query time largely depends on the implementation, especially for graph-based indexes. While a graph is being explored, the same nodes should not be evaluated repeatedly. Therefore, visited nodes should be managed during the search process. Since managing these nodes occupies a relatively large portion of query time, we also improve the managing of visited nodes to shorten the query time. In addition, we describe how to optimize our methods' parameters to attain the best search performance. In this paper, we make the following contributions.
\begin{itemize}
\item We propose the following adjustment methods for a graph. 
\begin{itemize}
\item Static degree adjustment for managing the degrees derived from a KNNG.
\item Static degree adjustment with constraints more precisely managing the degrees derived from a KNNG.
\item Dynamic degree adjustment, which depends on the required search accuracy during the search process.
\item Path adjustment for taking into account alternative paths in a graph.
\end{itemize}
We also show how to obtain the optimal parameters for static degree adjustments.
\item We improve the managing of nodes visited during the search process to shorten the query time.
\end{itemize}

\section{Proposed Methods}
Let $G = G(V, E)$ be a graph, where $V$ is a set of nodes that are objects in a $d$-dimensional vector space $\mathbb{R}^d$. The term $E$ is a set of directed edges, where an edge $e = \{u, v\}$ connects node $u$ to node $v$. In graph-based proximity searches, each node in a graph corresponds to an object to search for. In this paper, the graphs are neighborhood graphs in which neighboring nodes are associated with edges. Thus, neighboring nodes around any node can be directly obtained from the edges. Algorithm \ref{alg:knnsearch} shows our k-nearest neighbor search (KNN search) for obtaining k-nearest nodes to a query object from a neighborhood graph $G$. Let $q$, $k$, and $R$ be a query object, number of resultant objects, and a set of resultant objects, respectively. The term $\epsilon$ defines an exploration space $r_e = r(1 + \epsilon)$, where $r$ is a search radius. As $\epsilon$ increases, precision becomes higher while the query time increases. Therefore, precision and query time can be adjusted with $\epsilon$. The term $e_p$ is used for our dynamic degree adjustment explained in Section \ref{sec:da}. Let $e_p$ be $\infty$ so that our dynamic degree adjustment is not used. Let $C$ be a set of visited nodes, $d(x,y)$ be the distance between objects $x$ and $y$, and $N(G, s) $ be a set of neighboring nodes associated with the edges of node $s$ in graph $G$, where $N(G, s)=\{v|\{s,v\} \in E\}$. Each edge $e=\{u, v\}$ of the graphs discussed in this paper has a length that is distance $d(u, v)$, and the edges of each node are sorted by length. Therefore, $d(x, s)$ in line 11 causes no actual distance computation. Seed nodes $S$ used as starting nodes for exploring a graph can be obtained from the function Seed. Although it is possible for seed nodes to be randomly sampled from nodes in a graph, our proposed methods use the tree-based index in the same way as NGT\footnote{https://github.com/yahoojapan/NGT} to efficiently reach nodes roughly neighboring a query object. The tree-based index of the NGT is based on the vp-tree. 

\begin{algorithm}
{\fontsize{8pt}{10pt}\selectfont
\caption{KnnSearch}
\label{alg:knnsearch}
\begin{algorithmic}[1]
\REQUIRE $G, q, k, \epsilon, e_p$
\ENSURE $R$
\STATE $S \leftarrow \mathrm{Seed}(G)$, $r \leftarrow \infty$, $R \leftarrow S$
\STATE $C \leftarrow \emptyset$
\COMMENT{$ b \leftarrow \{empty, \cdots \}, L \leftarrow \{\emptyset, \cdots \}$}
\WHILE{$S \neq \emptyset$}
 \STATE $s \leftarrow \argmin_{x \in S} d(x, q)$, $S \leftarrow S \setminus \{s\}$
 \IF {$d(s, q) > r(1 + \epsilon)$}
 \RETURN $R$
 \ENDIF
 \STATE $p \leftarrow 1$
 \STATE $M \leftarrow N(G, s)$
 \WHILE {$M \neq \emptyset$ and $p \leq e_p$}
 \STATE $n \leftarrow \argmin_{x \in M} d(x, s)$
 \STATE $M \leftarrow M \setminus \{ n \} $ 
 \IF[{\bf if} $\mathrm{IsSet}(n)=false$ {\bf then}] {$n \notin C $}
 \STATE $C \leftarrow C \cup \{n\}$
 \COMMENT{$\mathrm{Set}(n)$\hspace{17mm}}
 \IF {$d(n, q) \leq r(1 + \epsilon)$}
 \STATE $S \leftarrow S \cup \{n\}$
 \ENDIF
 \IF {$d(n, q) \le r$}
 \STATE $R \leftarrow R \cup \{n\}$
 \IF {$|R| > k$}
 \STATE $R \leftarrow R \setminus \{\argmax_{x \in R} d(x, q)\}$
 \ENDIF
 \IF {$|R| = k$}
 \STATE $r \leftarrow \max_{x \in R} d(x, q)$
 \ENDIF
 \ENDIF
 \ENDIF
 \STATE $p \leftarrow p + 1$
 \ENDWHILE
\ENDWHILE
\RETURN $R$
\end{algorithmic} 
}
\end{algorithm}

\begin{figure}
\begin{center}
\vspace{0mm}
\hspace{-3mm}
\subfigure[] {
\includegraphics[height=3.2cm] {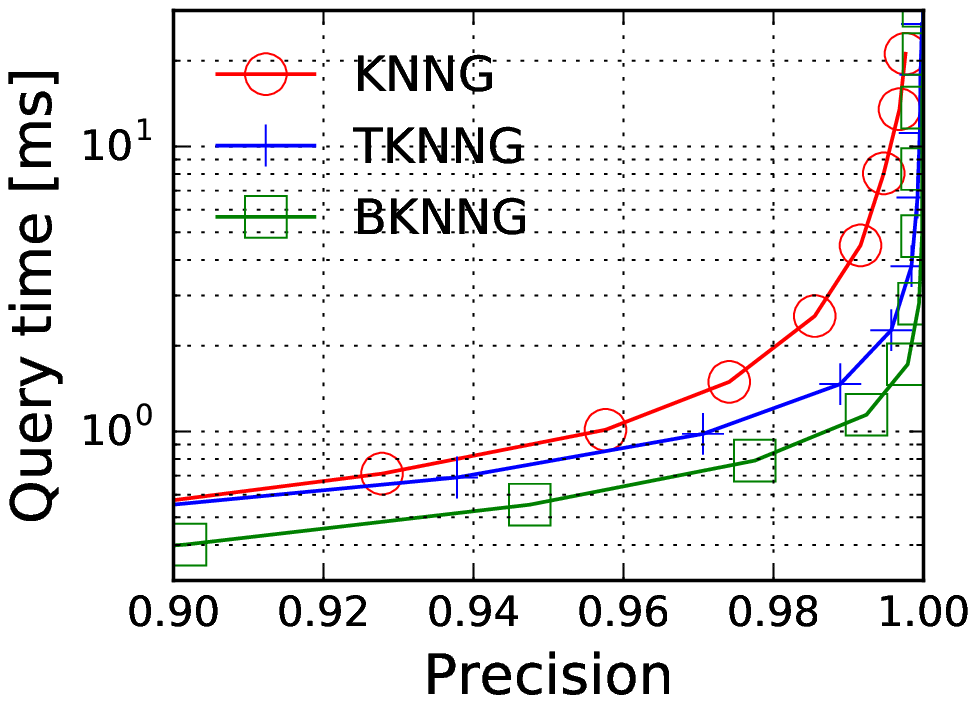}
}
\hspace{-4mm}
\subfigure[] {
\includegraphics[height=3.2cm] {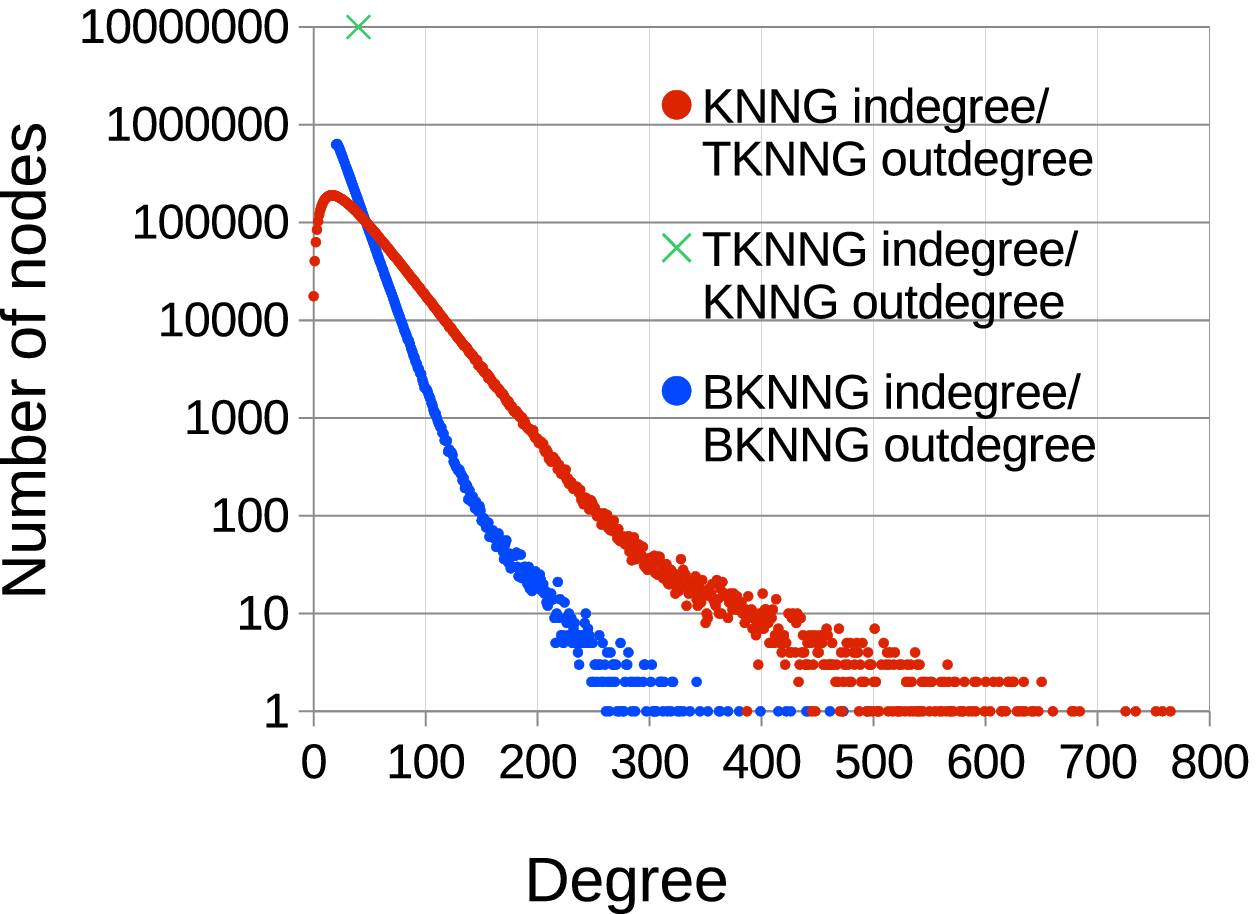}
}
\caption{(a) Query time vs. precision for 10 M SIFTs. (b) Frequency distributions of indegrees and outdegrees for all nodes.}
\label{fig:xnng-comp} 
\end{center}
\vspace{0mm}
\end{figure}

\begin{algorithm}
{\fontsize{8pt}{10pt}\selectfont
\caption{ConstructAdjustedGraph}
\label{alg:AdjustEdges}
\begin{algorithmic}[1]
\REQUIRE $G(V, E), e_o, e_i$
\ENSURE $G_e(V_e, E_e)$
\STATE $V_e \leftarrow V$, $E_e \leftarrow \emptyset$, $G_e \leftarrow (V_e, E_e)$
\FORALL{$o \in V$}
 \STATE $S \leftarrow N(G, o)$
 \STATE $p \leftarrow 1$
 \WHILE {$S \neq \emptyset$ and $p \leq e_o$ and $p \leq e_i$}
 \STATE $n \leftarrow \argmin_{x \in S} d(x, o) $
\STATE $S \leftarrow S \setminus \{ n \} $
\IF {$p \leq e_o$ and $N(G_e, o) \cap \{ n \} = \emptyset$}
 \STATE $N(G_e, o) \leftarrow N(G_e, o) \cup \{ n \}$
 \ENDIF
\IF {$p \leq e_i$ and $N(G_e, n) \cap \{ o \} = \emptyset$}
\STATE $N(G_e, n) \leftarrow N(G_e, n) \cup \{ o \}$
\ENDIF
\STATE $p \leftarrow p + 1$
\ENDWHILE
\ENDFOR
\RETURN $G_e$
\end{algorithmic}
}
\end{algorithm}

\begin{algorithm}
{\fontsize{8pt}{10pt}\selectfont
\caption{ConstructAdjustedGraphWithConstraint}
\label{alg:AdjustEdgesWithConstraint}
\begin{algorithmic}[1]
\REQUIRE $G(V, E), e_o, e_i$
\ENSURE $G_e(V_e, E_e)$
\STATE $G_t(V_t, E_t) \leftarrow $ConstructAdjustedGraph$(G, 0, e_i)$
\STATE $V_e \leftarrow V$, $E_e \leftarrow \emptyset$, $G_e \leftarrow (V_e, E_e)$
\STATE $V_i \leftarrow V$, $E_i \leftarrow \emptyset$, $G_i \leftarrow (V_i, E_i)$
\STATE $M \leftarrow V$
\WHILE {$M \neq \emptyset$}
 \STATE $o \leftarrow \argmin_{x \in M} \lvert N(G_t, x) \rvert$
 \STATE $M \leftarrow M \setminus \{ o \} $
 \STATE $S \leftarrow N(G_t, o)$
 \WHILE {$S \neq \emptyset$}
 \STATE $n \leftarrow \argmin_{x \in S} d(x, o) $
 \STATE $S \leftarrow S \setminus \{ n \} $
 \IF {$N(G_i, n) = \emptyset$ or $\{\lvert N(G_i, n)\rvert < e_i$ and $\lvert N(G_e, o) \rvert < e_o \}$}
 \STATE $N(G_e, o) \leftarrow N(G_e, o) \cup \{ n \}$
 \STATE $N(G_i, n) \leftarrow N(G_i, n) \cup \{ o \}$
 \ENDIF
 \ENDWHILE
\ENDWHILE
\FORALL{$o \in V$}
 \STATE $S \leftarrow N(G, o)$
 \WHILE {$S \neq \emptyset$ and $N(G_e, o) < e_o$}
 \STATE $n \leftarrow \argmin_{x \in S} d(x, o) $
 \STATE $S \leftarrow S \setminus \{ n \} $
 \IF {$N(G_e, o) \cap \{ n \} = \emptyset$}
 \STATE $N(G_e, o) \leftarrow N(G_e, o) \cup \{ n \}$
 \ENDIF
 \ENDWHILE 
\ENDFOR
\RETURN $G_e$
\end{algorithmic}
}
\end{algorithm}

\subsection{Static Degree Adjustment}
Fig.~\ref{fig:xnng-comp}(a) shows query time versus precision with Algorithm \ref{alg:knnsearch} for 10 M SIFT descriptors \cite{lowe1999object} for KNNG-based graph indexes. The curves were plotted by varying $\epsilon$. The KNNG's outdegrees were 40. The term TKNNG in the figure denotes a transposed KNNG derived by reversing all of the edge directions in the KNNG with outdegrees of 40. The bi-directed KNNG (BKNNG) was a KNNG with outdegrees of 20 to which was added the reversed edges of the KNNG with outdegrees of 20. Although the total number of directed edges for the BKNNG in the figure was almost the same as those of the KNNG and TKNNG, the query times were clearly different. Interestingly, the query time of the TKNNG was made shorter than that of the KNNG simply by reversing the edges. From this point, it is assumed that the reversed (incoming) edges of a KNNG for a TKNNG might be more effective in shortening the query time than the original edges of the KNNG. Moreover, the query time of the BKNNG was clearly shorter than that of the TKNNG. What brought about these trends?

To answer this, we examined the difference in performance in terms of indegree and outdegree. Fig.~\ref{fig:xnng-comp}(b) shows the frequency distributions of indegrees and outdegrees for all nodes in the KNNG, TKNNG, and BKNNG, which are the same graphs as in (a). Since the TKNNG is only a transposed KNNG, the distributions of outdegrees or indegrees for the KNNG are equal to those of the indegrees or outdegrees for the TKNNG. Since the BKNNG is a bi-directed graph, the distributions of indegrees and outdegrees in the BKNNG are completely the same. For the indegrees of the KNNG, there were more than 10,000 nodes of which indegrees were zero. Thus, these nodes  could not be reached during the search process. Moreover, since the indegrees of many of the nodes were less than 10, the probabilities of reaching them are very low. Thus, the search precision of the KNNG was reduced by nodes with such low indegrees. Since the indegrees for the TKNNG are constant, i.e., 40, the precision of the TKNNG was not reduced, unlike the KNNG. The outdegrees for the KNNG are constant, i.e., 40. The TKNNG had many nodes with outdegrees numbering several hundred, unlike the KNNG. For these nodes, to find the closest node neighboring a query, distances between all neighboring nodes and the query must be calculated. The presence of these excessive outdegrees is certain to increase the query time. The BKNNG did not have any nodes with an indegree of less than 20, and the outdegrees of most of the nodes in the BKNNG were less than that of the TKNNG. We therefore conclude that the query time of the BKNNG is shorter than those of the TKNNG and KNNG. 

From these observations, criteria for improving search performance are:
\begin{itemize}
\item High outdegrees that increase query time should be reduced.
\item Low indegrees that reduce precision should be increased.
\end{itemize}
An adjusted graph, in which the indegrees and outdegrees for each node are adjusted according to the criteria, can be derived from a KNNG as follows. Let $e_o$ and $e_i$ be the expected outdegrees and expected indegrees. First, an adjusted graph that has all objects as nodes without any edges is generated. A specified number $e_o$ of edges is extracted from each node in a KNNG in ascending order of length to adjust the outdegrees. Another specified number $e_i$ of edges is extracted in the same way to adjust the indegrees. The former edges are added to the nodes in the adjusted graph just as they were originally. The latter edges are reversed and added to the nodes in the adjusted graph as reverse (incoming) edges. Thus, the indegrees and outdegrees of the adjusted graph can be adjusted by varying the $e_o$ and $e_i$. Algorithm \ref{alg:AdjustEdges} shows exactly how to construct an adjusted graph $G_e$. 

\subsection{Static Degree Adjustment with Constraints}
With static degree adjustment, outdegrees and indegrees can be adjusted to some extent. However, since the original edge for a source node is the reverse edge for its destination node, adjusting the number of original and reverse edges in Algorithm \ref{alg:AdjustEdges} does not exactly adjust the outdegrees and indegrees. In fact, even when the number of reverse edges is set to a small number, some of the nodes tend to have high outdegrees. Therefore, we propose a method that adjusts the indegrees and outdegrees so as not to increase the outdegrees. The pseudo code is shown in Algorithm \ref{alg:AdjustEdgesWithConstraint}. First, TKNNG $G_t$ is constructed from a KNNG with ConstructAdjustedGraph. Let a KNNG, which had all its edges removed, be an initial adjusted graph $G_e$. Let $G_i$ be a graph that is a transposed $G_e$ only to obtain the indegrees of $G_e$ in this process. One node is selected in ascending order of the outdegrees of the nodes in the TKNNG. For each one of the neighboring nodes for the node selected in ascending order of edge length, if the indegree of a neighboring node in the $G_i$ is lower than $e_i$ and the outdegree of the node in the $G_e$ is lower than $e_o$, the edge to the neighboring node is added to the same node in the adjusted graph $G_e$. This is processed repeatedly until all of the nodes have been processed. Next, if the outdegrees of the nodes in the adjusted graph $G_e$ are less than $e_o$, the original edges are added to the nodes until the nodes have $e_o$ edges. Therefore, the indegrees and outdegrees can be more precisely adjusted.

\begin{figure}
\begin{center}
\includegraphics[width=0.9\linewidth] {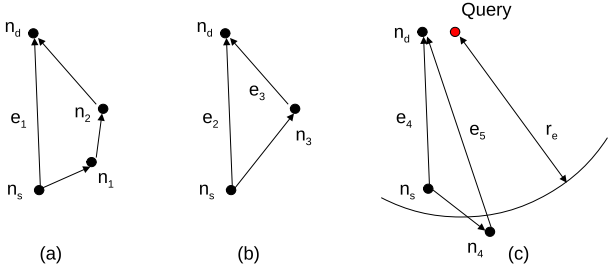}
\caption{Path adjustment}
\label{fig:edge-reduction}
\end{center}
\end{figure}

\begin{algorithm}
{\fontsize{8pt}{10pt}\selectfont
\caption{AdjustPath}
\label{alg:pruneedge}
\begin{algorithmic}[1]
\REQUIRE $G(V, E)$
\ENSURE $G_p(V_p, E_p)$
\STATE $V_p \leftarrow V$, $E_p \leftarrow \emptyset$
\STATE $G_t(V_t, E_t) \leftarrow G(V, E)$
\WHILE {$V_t \neq \emptyset$}
\FORALL{$n \in V_t$}
 \STATE $n_d \leftarrow \argmin_{x \in N(G_t, n)} d(n, x)$
\IF {$\mathrm{HasPath}(G_p, n, n_d) = false$}
\STATE $N(G_p, n) \leftarrow N(G_p, n) \cup \{ n_d\}$
\ENDIF
\STATE $N(G_t, n) \leftarrow N(G_t, n) \setminus \{ n_d\}$
\IF {$N(G_t, n) = \emptyset$}
 \STATE $V_t \leftarrow V_t \setminus \{ n\}$
\ENDIF
\ENDFOR
\ENDWHILE
\RETURN $G_p$
\end{algorithmic}
}
\end{algorithm}

\begin{algorithm}
{\fontsize{8pt}{10pt}\selectfont
\caption{HasPath}
\label{alg:AlternativePath}
\begin{algorithmic}[1]
\REQUIRE $G, n_s, n_d$
\ENSURE Whether there is an alternative path: $true$ or $false$
\FORALL{$n \in N(G, n_s)$}
\IF {$N(G, n) \cap \{ n_d \} \not= \emptyset $ and $d(n, n_d) < d(n_s, n_d)$}
 \RETURN $true$
\ENDIF
\ENDFOR
\RETURN $false$
\end{algorithmic}
}
\end{algorithm}

\subsection{Path Adjustment}
The above two static-degree adjustments take into account only optimizing the relationships between two objects connected by edges. The paths that are traversed during searches should be optimized. There are many long shortcut edges that can be substituted with an alternative path. Such edges can be effective in skipping nodes to reach the nodes further from a seed node. However, since our proposal uses a tree-based index to obtain nodes roughly neighboring a query as seed nodes, these shortcut edges are unnecessary. Thus, removing such edges can reduce the outdegrees to shorten query time. Even if edges can be removed, the query time will not necessarily decrease. If an alternative path consists of several nodes, the query time will increase because the number of distance computations increases to traverse these nodes on the path during the search process. Our path adjustment removes only edges that can be replaced with an alternative path that consists of two edges. In the case shown in Fig.~\ref{fig:edge-reduction}(a), target edge $e_1$ is not removed because its alternative path $\{n_s, n_1, n_2, n_d\}$ consists of three edges. A target edge means an edge that is checked to determine whether it should be removed. In the case shown in (b), target edge $e_2$ should be removed because its alternative path $\{n_s, n_3, n_d\}$ consists of only two edges. However, even though an alternative path consists of only two edges, if any of these edges on the path is longer than the target edge, the target edge is not removed. This case is shown in Fig.~\ref{fig:edge-reduction}(c), where edge $e_5$ on the alternative path of target edge $e_4$ is longer than the target edge. During the search process, since node $n_4$ on the alternative path with the long edge might be outside the search range, as Fig.~\ref{fig:edge-reduction}(c) shows, such an alternative path cannot be traversed according to Algorithm \ref{alg:knnsearch}. Therefore, edge $e_4$ should not be removed. Algorithm \ref{alg:pruneedge} precisely shows how to adjust paths. HasPath$(G, n_s, n_d)$ returns whether an edge from $n_s$ to $n_d$ has an alternative path. HasPath is shown in Algorithm \ref{alg:AlternativePath}. Although the PANNG also prunes long shortcut edges, it does not take into account the distance between a shortcut edge and the edge of an alternative path, as Fig.~\ref{fig:edge-reduction}(c) shows. Moreover, although our method removes all of the shortcut edges in a graph, the PANNG removes shortcut edges only for the long edges of each node. Therefore, the effectiveness of the PANNG in pruning edges is limited.

\subsection{Dynamic Degree Adjustment}
\label{sec:da}
Our static degree adjustment and path adjustment can almost adjust the outdegrees and indegrees with specified expected degrees to construct a static graph. However, if high precision is required, a high outdegree is indispensable. If a short query time is prioritized over high accuracy, a high outdegree, which increases the query time, is unnecessary. Therefore, outdegrees should be dynamically adjusted on the basis of a user's required search accuracy. The number of edges explored during the search process should be determined by the required precision at the beginning of the search. Search precision depends on the $\epsilon$ for the search process. Thus, the number $e_p$ of explored edges should be determined by using a specified $\epsilon$. We defined $e_p$ with the following formula.
\begin{equation}
e_p = 10^{w_e\epsilon}+e_0
\end{equation}
The minimum outdegree is $1 + e_0$, where the minimum $\epsilon = 0.0$, and the increase rate is defined by $w_e$ for $\epsilon$. The pseudo search code with this $e_p$ was already shown in Algorithm \ref{alg:knnsearch}.

\begin{algorithm}
{\fontsize{8pt}{10pt}\selectfont
\caption{Set}
\label{alg:set}
\begin{algorithmic}[1]
\REQUIRE $n_i$
\STATE $h = i \bmod s$
\IF {$b_h =$ empty}
 \STATE $b_h = i$
\ELSE
 \IF {$b_h \neq i$}
 \STATE $L_h \leftarrow L_h \cup \{ i\}$
 \ENDIF
\ENDIF
\RETURN
\end{algorithmic}
}
\end{algorithm}

\begin{algorithm}
{\fontsize{8pt}{10pt}\selectfont
\caption{IsSet}
\label{alg:isSet}
\begin{algorithmic}[1]
\REQUIRE $n_i$
\ENSURE Whether $n_i$ has been visited: $true$ or $false$
\STATE $h = i \bmod s$
\IF {$b_h = i$}
 \RETURN $true$
\ENDIF
\IF {$L_h \cap \{i \} = \emptyset $}
 \RETURN $false$
\ENDIF
\RETURN $true$
\end{algorithmic}
}
\end{algorithm}

\subsection{Improving Visited-Node Management}
For search processing, the time taken to manage what nodes have already been visited in a graph occupies a relatively large proportion of query time, especially to obtain a high precision. The most straightforward method is to use an array, where each entry that corresponds to all nodes in a graph shows whether a node has been visited. However, it takes a long time to initialize an array at the beginning of exploring a graph, especially when a large number of objects are stored in the graph. Thus, associative containers using a hash table provided by, for example, the C++ Standard Library or Boost\footnote{http://www.boost.org/}, are generally used to maintain visited nodes without being affected by the number of stored nodes. However, visited nodes are only a small proportion of all nodes, and they are checked many times regardless of whether they have already been visited while a graph is being explored. The generic hash tables provided by the C++ Standard Library and Boost are not fast enough for this specific case. A Bloom filter is also not fast enough for this case because using multiple hashes increases the checking time. Thus, we customized the hash table for this case. First, to avoid hash collisions and reduce the initialization cost, the minimum required size of the hash table was determined. Second, a bit operation was adopted for the hash function of the table to reduce the time taken to calculate hash values. Third, the first inserted objects for each hash value are stored in a table to accelerate checking. The second and following objects causing collisions are stored as a list. Algorithm \ref{alg:set}, Set, is a function that inserts a visited node into the visited node set $C$ in Algorithm \ref{alg:knnsearch}. Algorithm \ref{alg:isSet}, IsSet, returns whether a specified node exists in $C$. These functions respectively correspond to lines 14 and 13 in Algorithm \ref{alg:knnsearch}, where the revised source code is described in the comments. Let $n_i$ be a node in a set of nodes $V=\{n_1, n_2, \cdots, n_m \}$ in a graph, where $m$ is the number of all nodes. The computation in line 1 of these algorithms is a hash function, where $s$ is the hash table size. Let the hash table be $\{b_1, b_2, \cdots, b_s\}$, and let $L_h$ be a bucket for a hash value $h$ in the hash table. Since $s$ should be a power of 2, the hash function in line 1 can be represented by only one bitmask operation. Since $s$ should depend on the number of objects in a graph, it is calculated with the formula 
\[
s=2^{\lfloor (\log_2 n + b) / 2 \rfloor}, 
\]
where $b$ determines the minimum hash size, which is $b=11$ in this paper. This formula and $b$ were derived from our preliminary experiments to determine the optimal size of the hash table to shorten the query time. $\log_2$ was adopted because it can be calculated faster than $\log$ with a bit operation. The calculation cost in line 2 in Algorithm \ref{alg:isSet} is significantly smaller than that in line 5. In addition, since most cases satisfy the condition in line 2, the processing cost of IsSet is extremely small for these cases. 

\begin{algorithm}
{\fontsize{8pt}{10pt}\selectfont
\caption{ConstructGraph}
\label{alg:constructGraph}
\begin{algorithmic}[1]
\REQUIRE $G_a, k_c, e_o, e_i $
\ENSURE $G_{adj}$
\STATE $G_k \leftarrow \mathrm{ConstructAdjustedGraph}(G_a, k_c, 0)$
\STATE $G_e \leftarrow \mathrm{ConstructAdjustedGraph}(G_k, e_o, e_i)$ or \\
\hspace{6.5mm} $\mathrm{ConstructAdjustedGraphWithConstraint}(G_k, e_o, e_i)$
\STATE $G_{adj} \leftarrow \mathrm{AdjustPath}(G_e)$
\RETURN $G_{adj}$
\end{algorithmic}
}
\end{algorithm}

\begin{figure*}
\begin{center}
\vspace{0mm}
\subfigure[] {
\includegraphics[width=4.5cm] {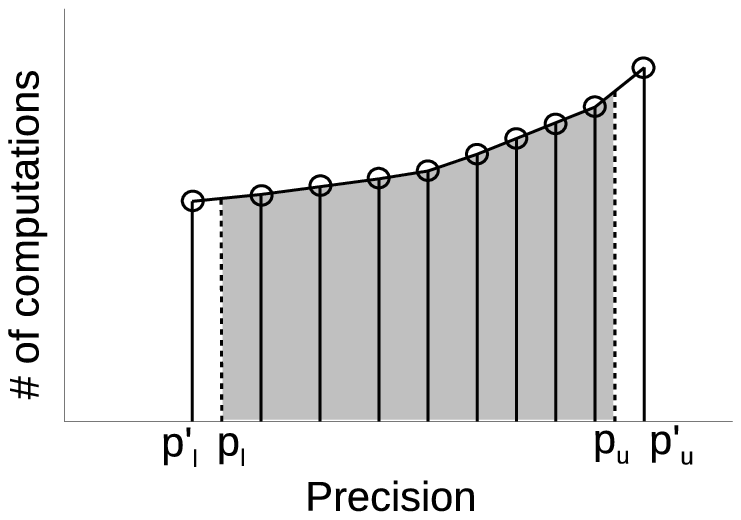}
}
\hspace{4mm}
\subfigure[] {
\includegraphics[width=4.5cm] {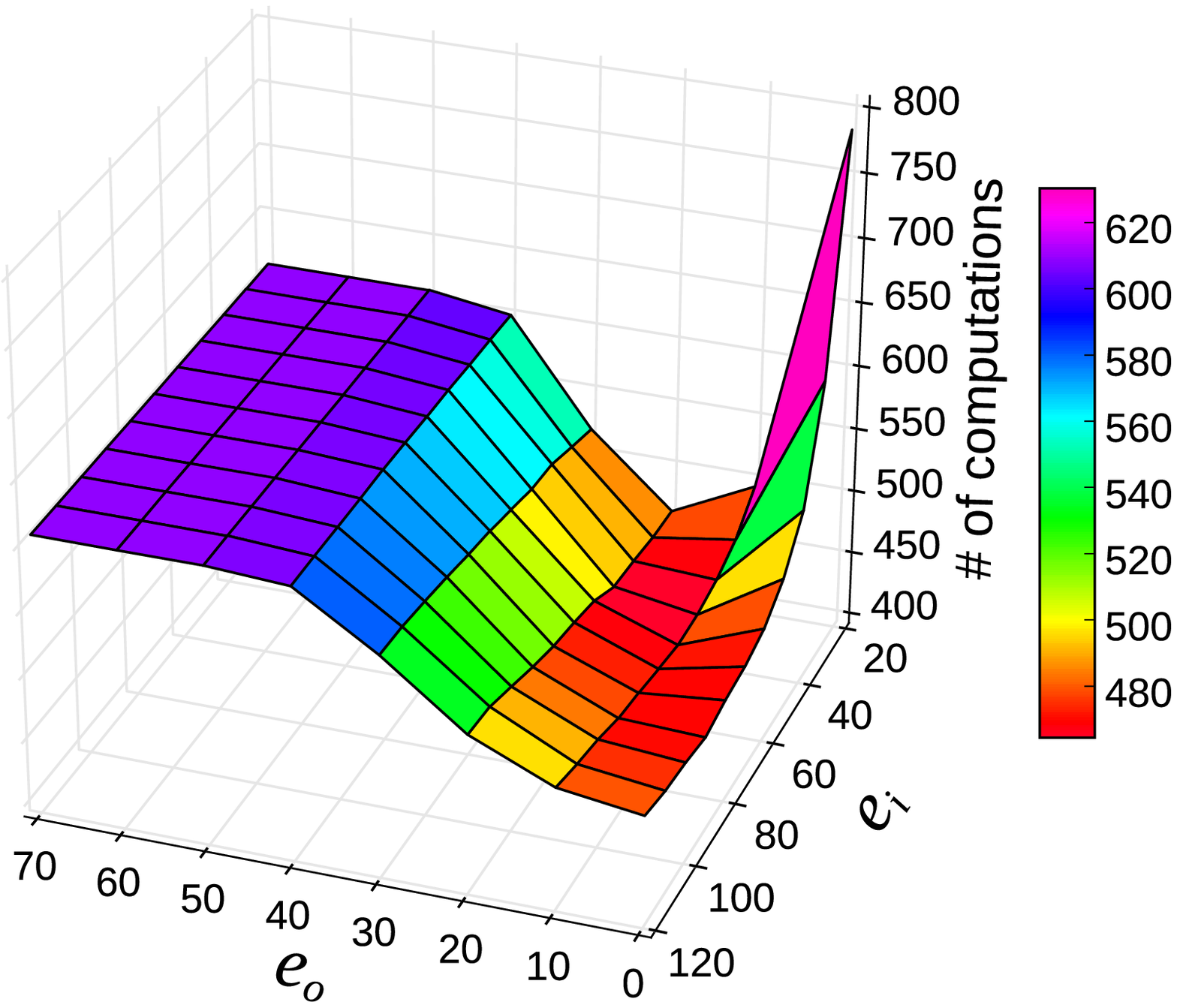}
}
\hspace{0mm}
\subfigure[] {
\includegraphics[height=3.2cm] {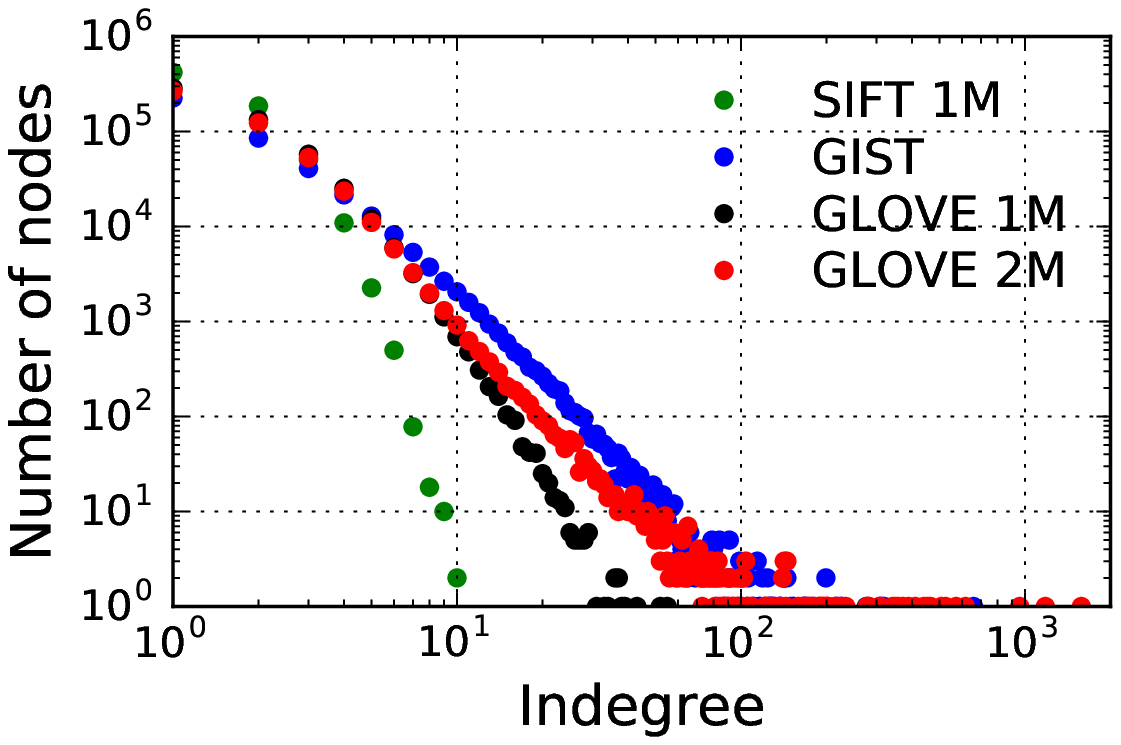}
\hspace{2mm}
}
\caption{(a) Method of calculating mean number of distance computations. (b) Mean number of distance computations vs. expected degree parameters $e_i$ and $e_o$ with DA for 1 M SIFTs. (c) Indegree histograms of 1-NNG. }
\label{fig:degree-optimization}
\end{center}
\vspace{0mm}
\end{figure*}

\subsection{Graph Construction and Optimization}
Algorithm \ref{alg:constructGraph} shows the complete procedure for constructing an adjusted graph $G_{adj}$. Let $G_a$, $G_k$, and $G_e$ be the ANNG, approximate KNNG (AKNNG), and degree-adjusted graph, respectively. Since the cost of constructing a KNNG is generally huge, we instead use an AKNNG that is derived from an ANNG. Appendix shows the ANNG construction with the pseudo code. Algorithm \ref{alg:constructGraph} shows the complete procedure to construct an adjusted graph $G_{adj}$. Let $k_c$ be the number of edges in the AKNNG with $k_c > e_o$ and $k_c > e_i$.

To construct optimized graphs, only optimal expected degree parameters $e_i$ and $e_o$ have to be determined. Thus, a loss function has to be defined to optimize them. Search performance can generally be measured with the precision and number of distance computations or query time. The actual query time is unstable for common multi-core systems. Moreover, the query time for each method depends on its implementation. Therefore, we use a stable number of distance computations instead of query time for our optimization. Moreover, we define a specific target precision range that is specified in advance because it is impossible to optimize a graph that always produces the best search performance for a wide precision range. Therefore, we define the mean number of distance computations for the target precision range as the loss function. The loss function $L(e_o, e_i, p_l, p_u)$ for the target precision range $[p_l, p_u]$ is calculated as
\begin{equation}
L(e_o, e_i, p_l, p_u)=\int_{p_l}^{p_u} \log_{10} C(e_o, e_i, x) dx /(p_u-p_l). 
\end{equation}
Let $C(e_o, e_i, p)$ be a function that returns the number of distance computations, where its precision is $p$. Since $C(e_o, e_i, p)$ exponentially increases for a high $p$, $\log_{10}$ is applied to $C(e_o, e_i, p)$ to suppress the effect on the high precision range. The integration value $\int_{p_l}^{p_u} \log_{10} C(e_o, e_i, x) dx$ is calculated with numerical integration by using the trapezoidal rule from samples that are produced from actual searches, as shown in Fig.~\ref{fig:degree-optimization}(a). First, $\epsilon_l$ and $\epsilon_u$, which output $p_l^{\prime}$ and $p_u^{\prime}$, where $p_l-0.005 < p_l^{\prime} \leq p_l$ and $p_u \leq p_u^{\prime} < p_l+0.005$, respectively, are determined by executing the search process with a binary search. Second, [$\epsilon_l$, $\epsilon_u$] is divided equally into 9 partitions to obtain 10 values of $\epsilon$ including $\epsilon_l$ and $\epsilon_u$. Finally, the integration value $\int_{p_l}^{p_u} \log_{10} C(e_o, e_i, x) dx$ is calculated from these 10 values of $p$ and $C(e_o, e_i, p)$, which are obtained by executing the search process with these 10 values of $\epsilon$, as Fig.~\ref{fig:degree-optimization}(a) shows. Fig.~\ref{fig:degree-optimization}(b) shows the mean numbers of distance computations $10^{L(e_o, e_i, 0.90, 0.98)}$ versus combinations of the $e_o$ and $e_i$ of a graph, which were constructed with Algorithm \ref{alg:constructGraph}. Since $L(e_o, e_i, p_l, p_u)$ is a convex function as seen in the figure, the minimum value can be found by using a simple hill-climbing algorithm. 

\section{Experimental Results}
We used SIFT 1M, GIST, and SIFT 10M from the TEXMEX dataset\footnote{http://corpus-texmex.irisa.fr/} and GLOVE 1M and 2M from the GLoVe\footnote{https://nlp.stanford.edu/projects/glove/} dataset for experiments. SIFT 1M is an ANN\_SIFT1M that consists of 1 M 128-dimensional SIFT local image descriptors. GIST is the ANN\_GIST1M that consists of 1 M 960-dimensional GIST global image descriptors \cite{oliva2001modeling}. SIFT 10M consists of 10 M SIFT descriptors that were randomly sampled from ANN\_SIFT1B. GLOVE 2M consists of 2,095,017 300-dimensional pre-trained word vectors that were generated from 840B tokens, and GLOVE 1M consists of 1,092,514 100-dimensional pre-trained word vectors generated from 27B tokens of Twitter. Each dataset contained 1,000 queries and 100,000 training objects. The training objects were used to optimize graphs. Each object was stored in memory as a 4-byte floating point number. The number of resultant objects was set to 20. The Euclidean distance function was used for the SIFTs and GIST, and the angular distance was used for the GLOVEs. We conducted the experiments on a computer with Intel Xeon E5-2630L (2.0 GHz and 64 GB of memory) CPUs. Although the CPUs had multiple cores, the experimental software was not run in parallel for the search process.

Instead of a KNNG, we constructed all KNNG-based indexes from an ANNG, which was constructed by using the NGT, where the number of edges $k_c$ was 200 and its construction parameter $\epsilon_c$ was a recommended value of $0.1$. Since the search performance of adjusted graphs derived from AKNNGs depends on the accuracies of the original AKNNGs, one thousand nodes sampled from each AKNNG and neighboring nodes associated with edges were evaluated. Table \ref{tbl:aknng-accuracy} shows the edge accuracies of the AKNNGs constructed from all of the datasets. Precision is the mean precision for the edges of each node in an AKNNG. The mean rank is the mean correct rank of all of the edges for each node. Therefore, if the mean precision of a node is 1.0, the mean rank of the node is 100.5, which is a mean from 1 to 200.

\begin{table}
\begin{center}
\caption{Edge precision and mean rank of AKNNGs constructed by using NGT library for all datasets}
\label{tbl:aknng-accuracy}
\vspace{-3mm}
\scriptsize
\begin{tabular}{|c||c|c|c|c|c|}
\hline
Accuracy & SIFT & GIST & GLOVE & GLOVE & SIFT \\
       & 1M  &      & 1M    & 2M    & 10M \\
\hline\hline
Precision & 0.715 & 0.545 & 0.697 & 0.681 & 0.741 \\
\hline
Mean rank & 178.1 & 360.4 & 192.2 & 268.5 & 160.8 \\
\hline
\end{tabular}
\end{center}
\end{table}

\begin{table}
\begin{center}
\caption{Types of our adjustment methods}
\label{tbl:proposed-methods}
\vspace{-3mm}
\scriptsize
\begin{tabular}{|c||c|c|c|}
\hline
Type & Static degree & Path & Dynamic degree \\
& adjustment & adjustment & adjustment \\
\hline\hline
SA & No constraint & \checkmark & \\
SAC & Constraint & \checkmark & \\
DA & No constraint & \checkmark & \checkmark \\
\hline
\end{tabular}
\vspace{0mm}
\end{center}
\end{table}

\begin{table}
\begin{center}
\caption{Optimal parameters}
\label{tbl:optimized-edge}
\vspace{-3mm}
\scriptsize
\begin{tabular}{|c|c||c|c|c|c|c|}
\hline
Method & Parameter & SIFT & GIST & GLOVE & GLOVE & SIFT \\
 & & 1M & & 1M & 2M & 10M \\
\hline\hline
SA & $e_o$ & 30 & 160 & 130 & 200 & 50 \\
& $e_i$ & 10 & 5 & 10 & 0 & 20 \\
\hline
SAC & $e_o$ & 55 & 135 & 140 & - & 110 \\
& $e_i$ & 10 & 30 & 70 & - & 45 \\
\hline
DA & $e_o$ & 30 & 10 & 15 & 10 & 10 \\
& $e_i$ & 110 & 115 & 155 & 140 & 95 \\
\hline
\end{tabular}
\vspace{0mm}
\end{center}
\end{table}

\vspace{3pt}\noindent{\sl {\bfseries Dataset Characteristics.\hspace{3pt}}}
First, we clarify the characteristics of the datasets to help in analyzing our experimental results. For graph-based indexes, what reduces search performance is mainly the concentrations of objects in an object vector space. Generally, datasets originally have some concentration of objects. These concentrations are amplified by high dimensionality. When a graph-based index is constructed for the objects, these concentrations tend to make a distorted graph that seems to have a kind of black hole during the search process that reduces search performance. For better understanding of such distorted datasets, the frequency distributions of indegrees, which are indegree histograms, for 1-NNG for the 1M and 2M datasets are shown in Fig.~\ref{fig:degree-optimization} (c). Each of all of the datasets had only 1M objects for fair comparison. A node with a high indegree means that many other nodes are close to it, that is to say, it has many neighboring nodes. It is also assumed that objects with a high indegree are the center of these concentrations. Even though each node had one edge, the maximum indegree was 10 for SIFT because of such concentrations. However, since the indegree of SIFT was clearly lower than that of the other datasets from the figure, it is assumed that SIFT was not much more concentrated compared with the others. For GIST, since the maximum indegree was 659, it is assumed that GIST was more concentrated than SIFT, and this concentration was amplified by higher dimensionality than SIFT. Although the dimensionality of GLOVE 2M was lower than GIST, the maximum indegree was 1,869. It is assumed that the concentration came from the manner of generating GLOVE datasets, which is based on the frequency of words, because many words are associated with high-frequency words. The indegree of GLOVE 1M was much lower than that of GLOVE 2M because of the lower dimensionality, similar to the relationship between SIFT and GIST. These concentrated objects tend to have many edges, reaching the neighboring nodes to improve search precision during graph construction. However, such nodes with excessive edges cause a large number of distance computations, increasing the query time. 

\begin{figure*}
\begin{center}
\subfigure[SIFT 1M] 
{\includegraphics[height=3.2cm] {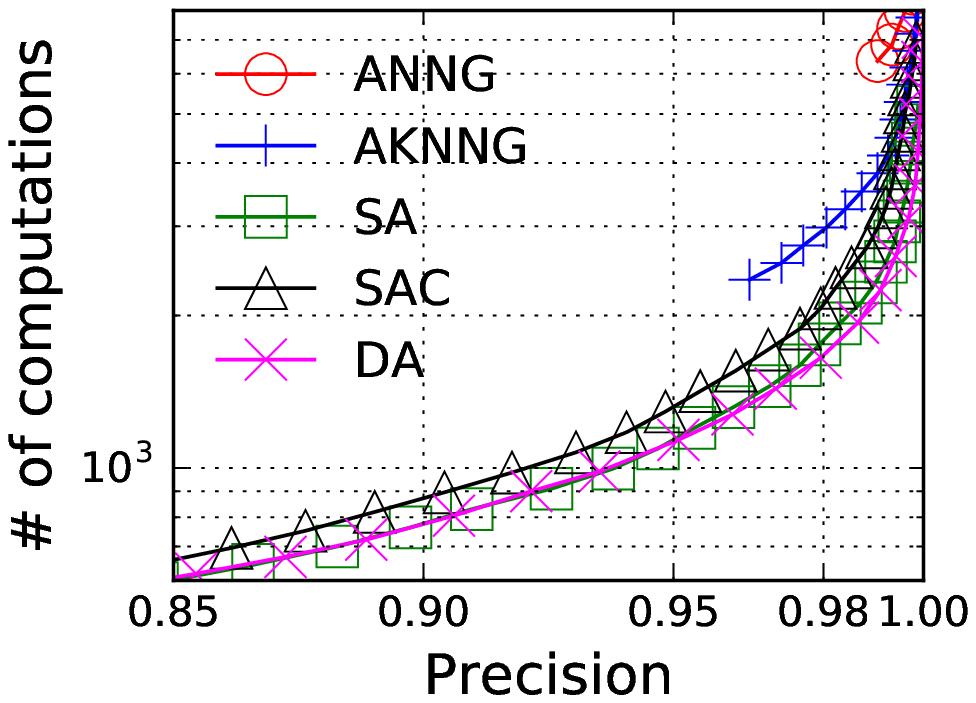}}
\hspace{0mm}
\subfigure[GIST] 
{\includegraphics[height=3.2cm] {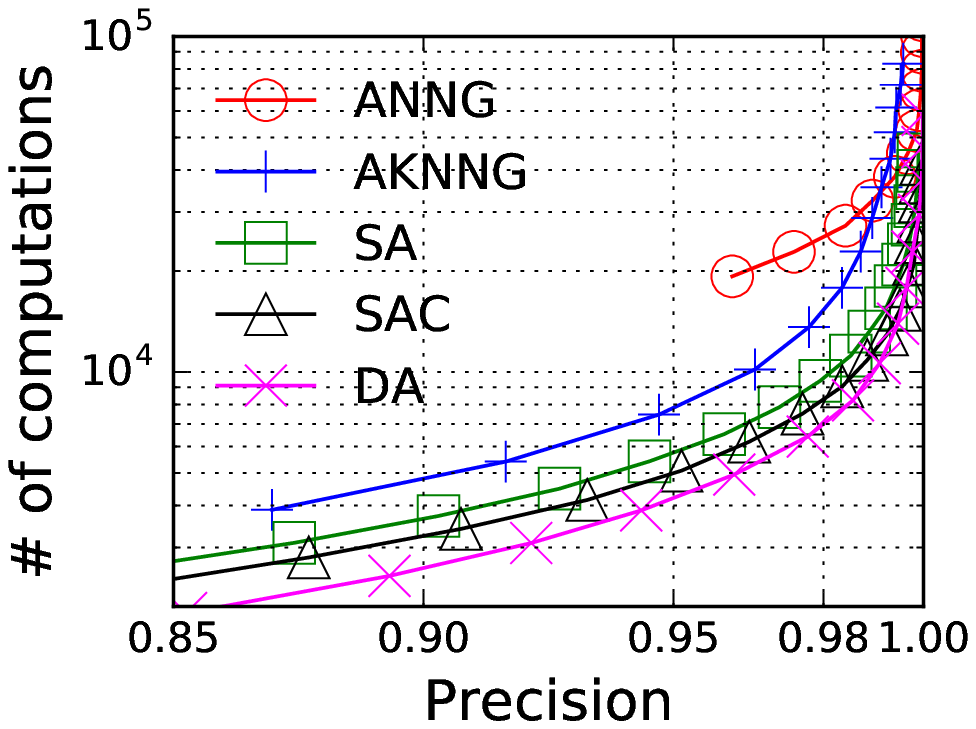}}
\hspace{0mm}
\subfigure[GLOVE 1M] 
{\includegraphics[height=3.2cm] {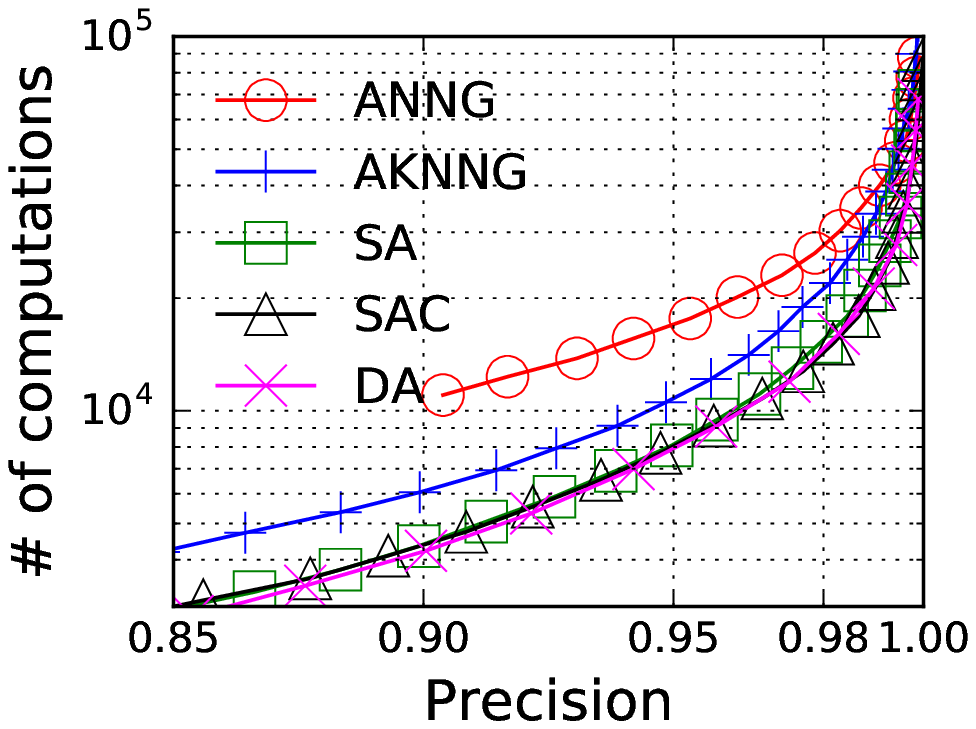}}
\hspace{0mm}
\subfigure[GLOVE 2M] 
{\includegraphics[height=3.2cm] {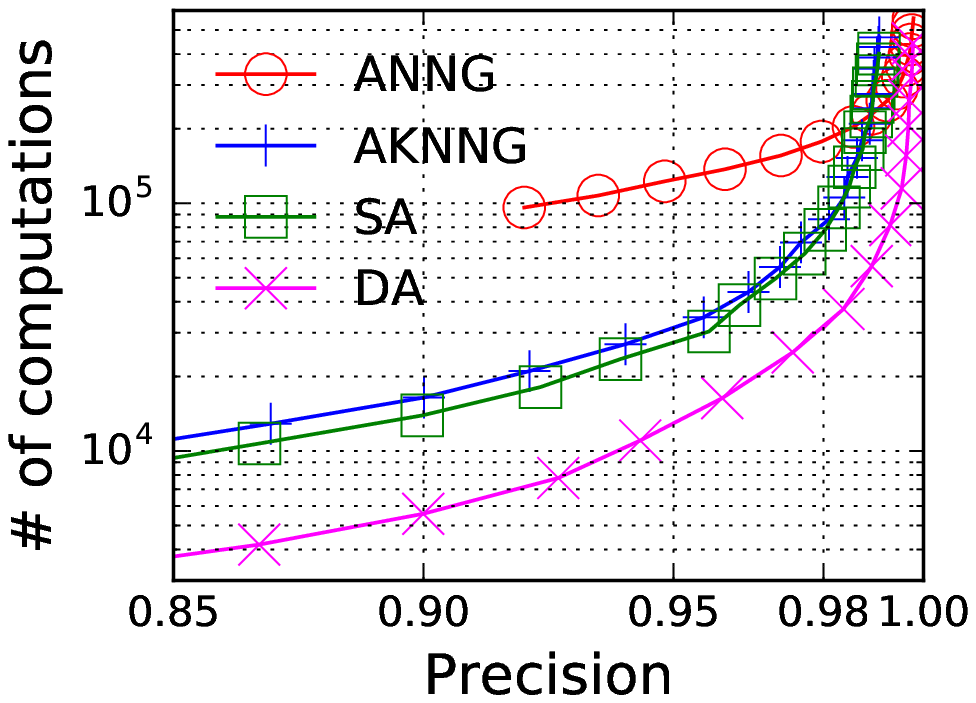}}
\caption{Number of distance computations vs. precision. Comparison among proposed methods.}
\label{fig:time-precision-1}
\vspace{0mm}
\end{center}
\end{figure*}

\begin{figure*}
\begin{center}
\subfigure[] {
\includegraphics[height=3.2cm] {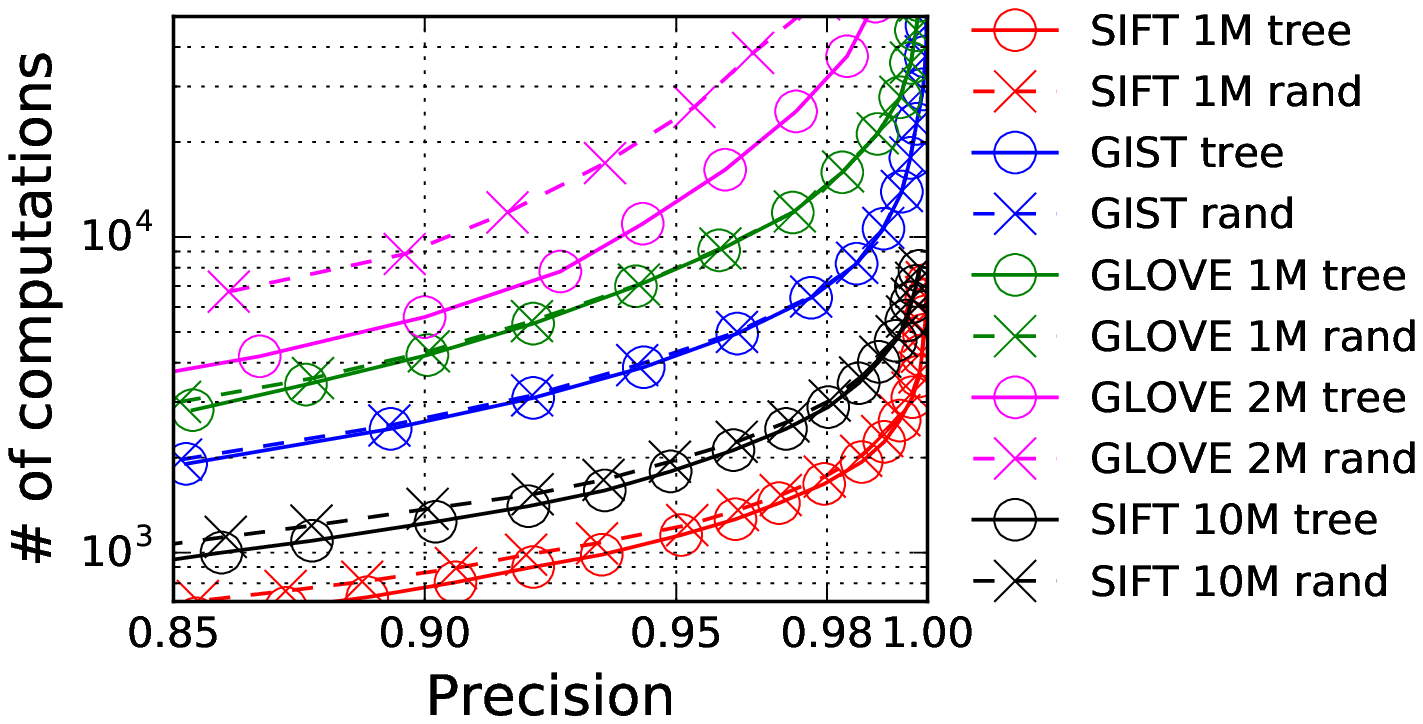}
}
\hspace{-2mm}
\subfigure[] {
\includegraphics[height=3.2cm] {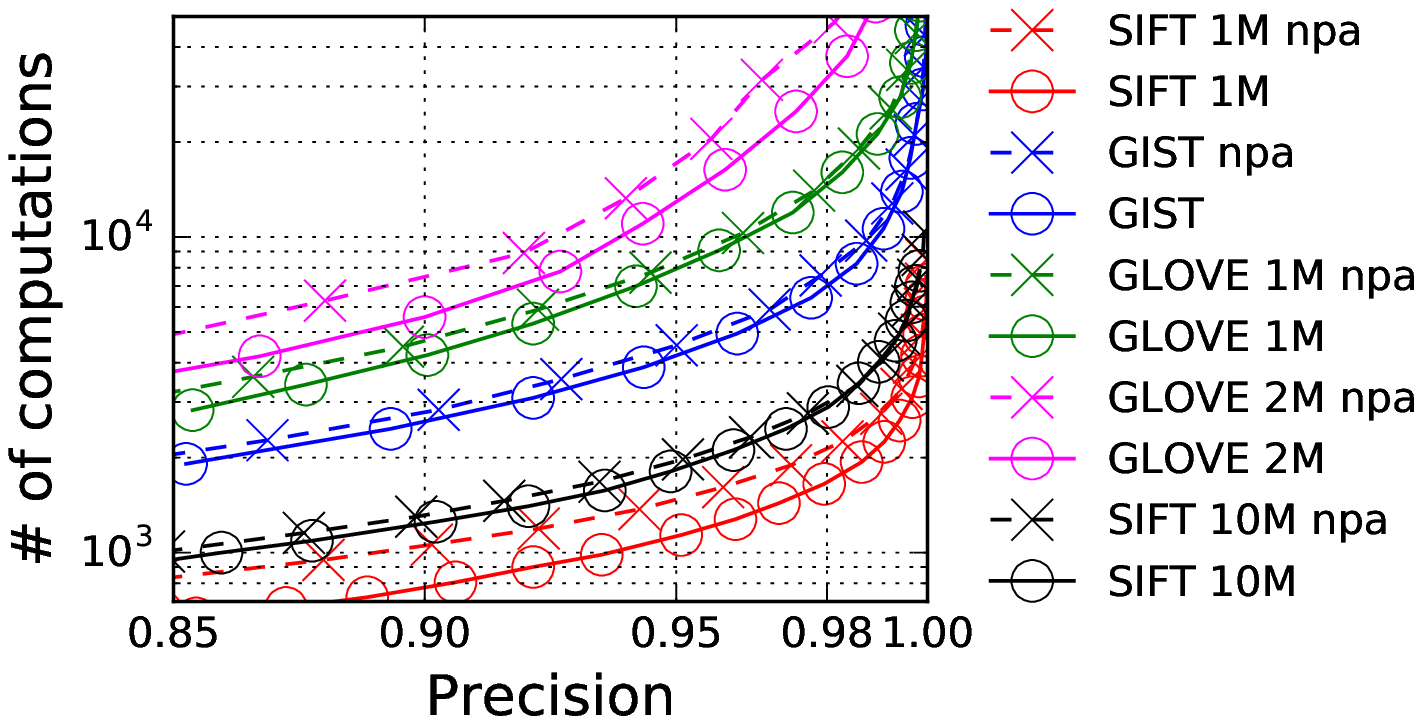}
}
\hspace{-2mm}
\subfigure[] {
\includegraphics[height=3.2cm] {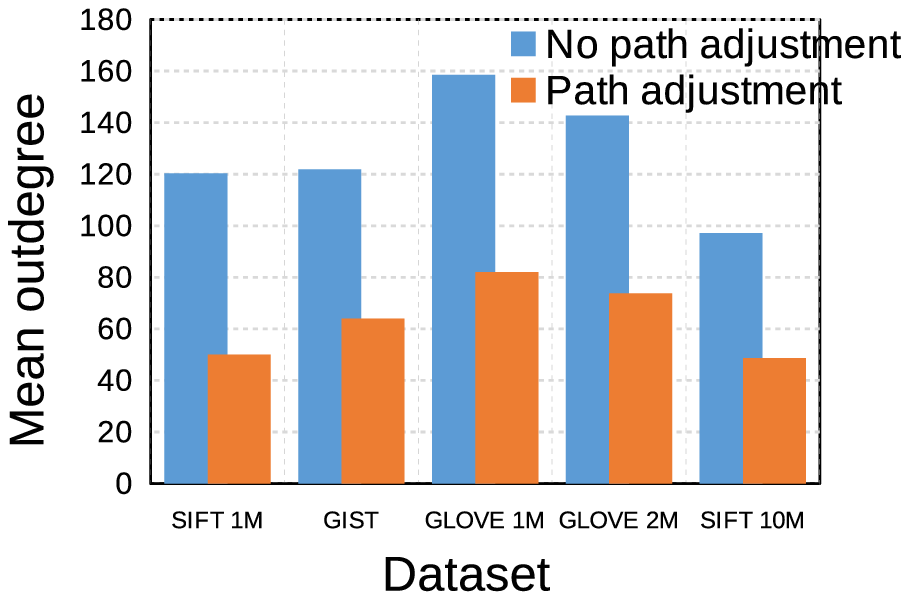}
}
\caption{(a) Number of distance computations vs. precision using tree-based index (``tree'') or random seeds (``rand''). (b) Number of distance computations vs. precision with and without using our path adjustment (``npa''). (c) Mean outdegree of each edge for all datasets with and without using our path adjustment.}
\label{fig:path-adjust-effect}
\vspace{0mm}
\end{center}
\end{figure*}

\begin{figure*}
\begin{center}
\subfigure[] {
\includegraphics[height=3.2cm] {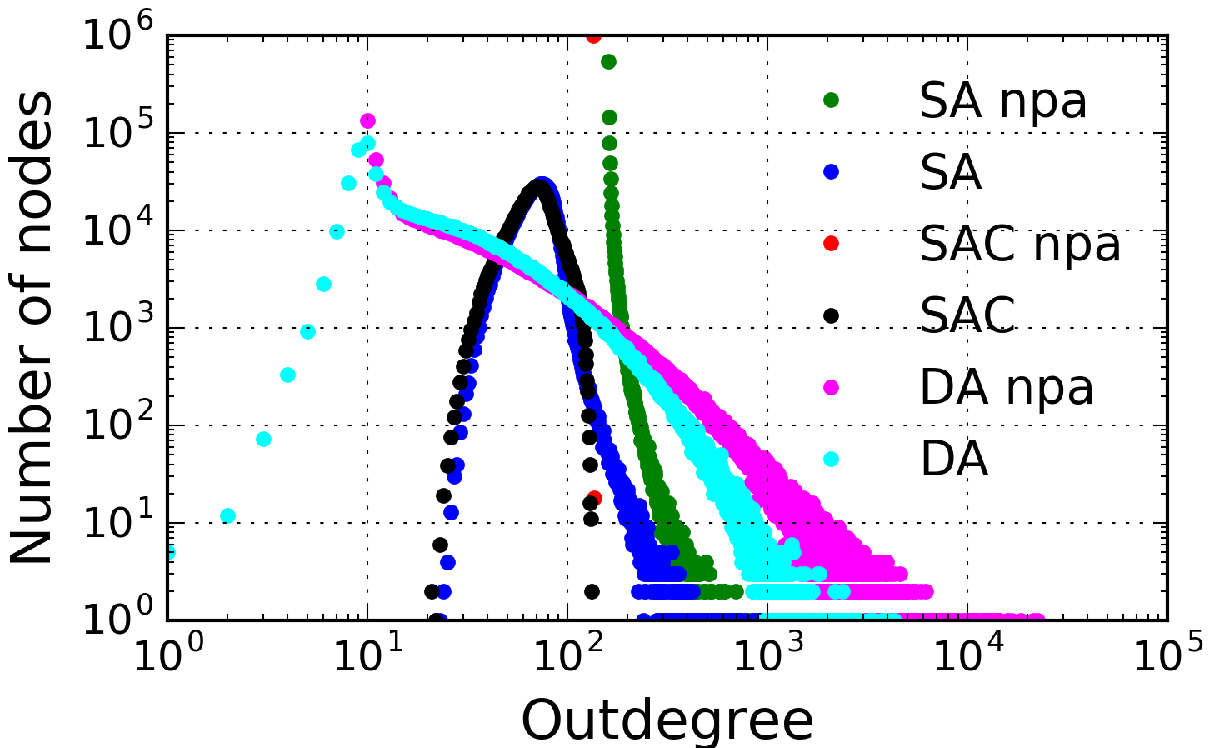}
}
\hspace{-2mm}
\subfigure[] {
\includegraphics[height=3.2cm] {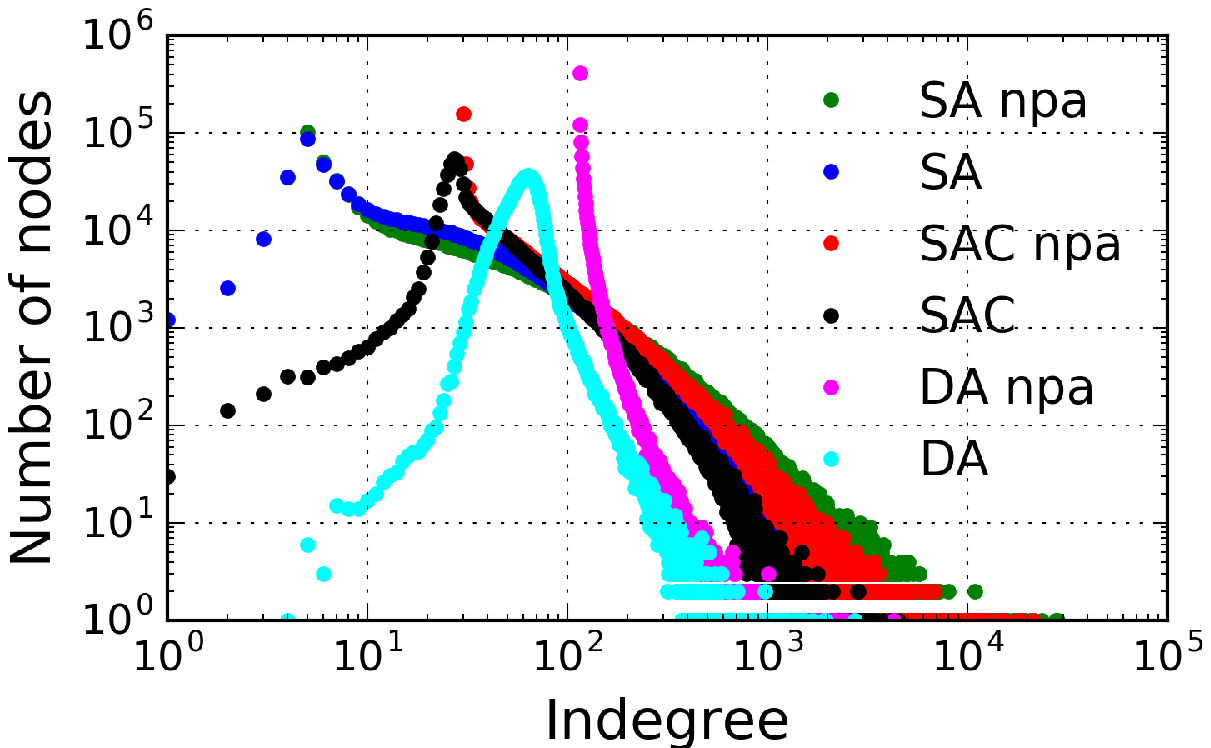}
}
\hspace{-2mm}
\subfigure[] {
\includegraphics[height=3.2cm] {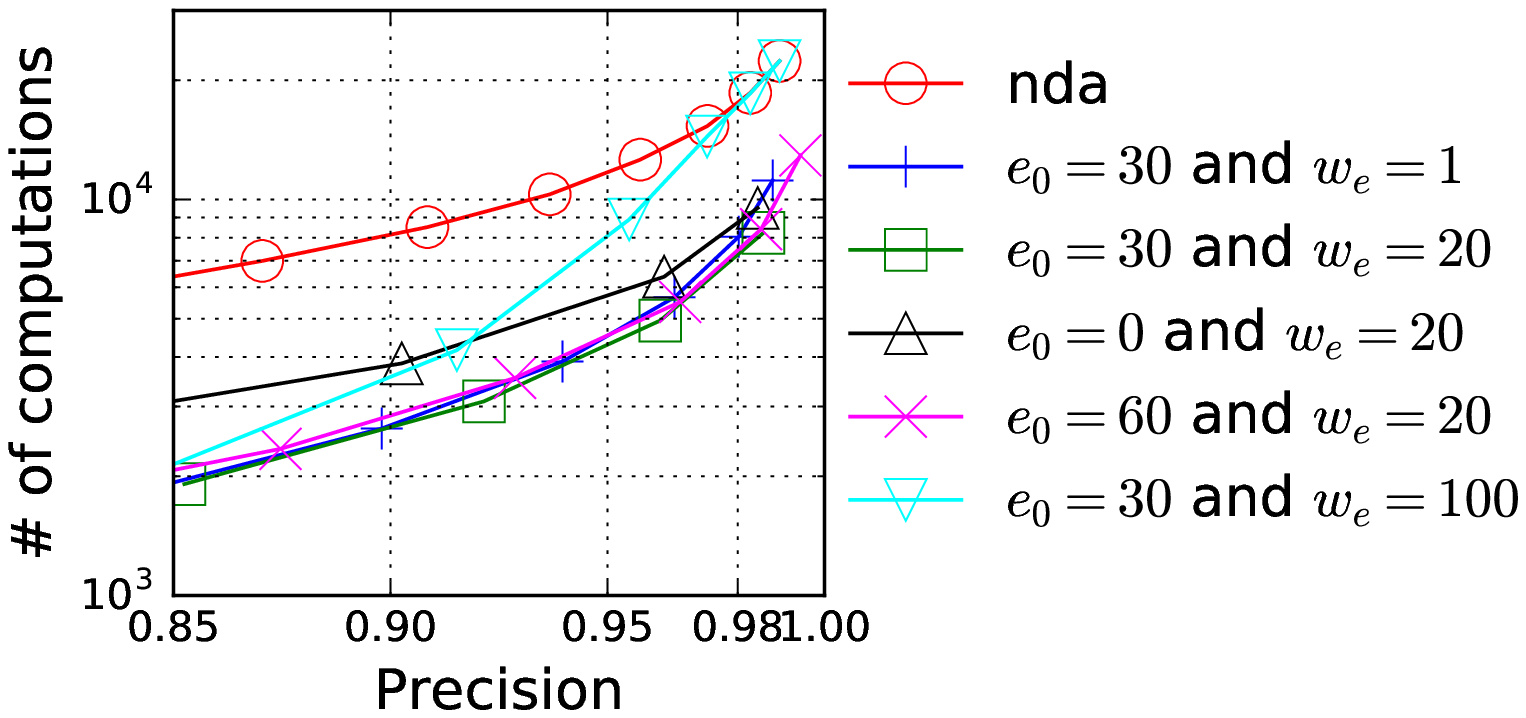}
}	
\caption{(a) Outdegree histogram of each node for GIST. (b) Indegree histogram of each node for GIST. (c) Number of distance computations vs. precision with varying dynamic degree-adjustment parameters $e_0$ and $w_e$ for GIST; ``nda'' represents not using our DA.}
\label{fig:dgree-histogram} 
\vspace{0mm}
\end{center}
\end{figure*}

\vspace{3pt}\noindent{\sl {\bfseries Graph-Degree Optimization.\hspace{3pt}}}
Since most applications generally require higher precision, we focused on reducing the number of distance computations for the high precision range from $p_l=0.90$ to $p_u=0.98$ for $L(e_o, e_i, p_l, p_u)$. The step of the hill climbing for the number of edges was set to 5. We now define our three types of degree adjustments for our experiments, SA, SAC, and DA, as Table \ref{tbl:proposed-methods} shows. Table \ref{tbl:optimized-edge} shows the expected degree parameters that were optimized for all datasets and types. The parameters of GLOVE 2M were not able to be obtained for SAC because the precision could not reach the target precision range due to its constraints. It is assumed that GLOVE 2M would be too distorted. 

\begin{table}
\begin{center}
\caption{Statistics of graphs constructed with our methods for all datasets}
\label{tbl:outdegree}
\vspace{-3mm}
\scriptsize
\begin{tabular}{|c|c||c|c|c|c|c|}
\hline
& Method & SIFT & GIST & GLOVE & GLOVE & SIFT \\
& & 1M & & 1M & 2M & 10M \\
\hline
\hline
Mean	&	SA	&	30.4 	&	109.1 	&	107.7 	&	162.7 	&	52.5 	\\
top 5\%	&	SAC	&	39.4 	&	110.4 	&	118.6 	&	-	&	76.5 	\\
outdegree	&	DA	&	120.7 	&	383.4 	&	287.6 	&	424.1 	&	109.9 	\\
\hline
Mean	&	SA	&	6.8 	&	3.7 	&	14.9 	&	1.5 	&	15.7 	\\
bottom 5\%	&	SAC	&	9.0 	&	19.0 	&	37.6 	&	-	&	30.0 	\\
indegree	&	DA	&	32.6 	&	37.0 	&	41.3 	&	32.3 	&	31.8 	\\
\hline
Mean	&	SA	&	206.5 	&	1.066 	&	0.9044 	&	0.9836 	&	221.4 	\\
indegree	&	SAC	&	212.0 	&	1.145 	&	0.9099 	&	-	&	223.1 	\\
distance	&	DA	&	207.5 	&	1.119 	&	0.9037 	&	0.9801 	&	221.5 	\\\hline
\end{tabular}
\vspace{0mm}
\end{center}
\end{table}

\subsection{Comparison among Our Proposed Methods}
Fig.~\ref{fig:time-precision-1} shows the number of distance computations versus precision with the ANNG, AKNNG, and our methods with the optimal expected degree parameters in Table \ref{tbl:optimized-edge} for the 1M and 2M datasets. Our SA, SAC, and DA were derived from the ANNG and AKNNG in Fig.~\ref{fig:time-precision-1}. The figure shows that DA always required the smallest number of distance computations. However, the curves of SA for SIFT 1M and GLOVE 1M were close to those of DA. Since these datasets were not distorted much, as Fig.~\ref{fig:degree-optimization}(c) shows, it is assumed that even SA can construct almost the best graph structure over a target precision range without dynamic degree adjustment. For GIST, the number of computations of SAC was smaller than that of SA, unlike other datasets. SAC may be effective for high-dimensional datasets like GIST. Our DA was the most effective for high-dimensional objects among the three. In some cases, since the number of degrees for SA and SAC were lower than that of DA, if reduction of memory usage should be prioritized, SA or SAC is another option. 

\vspace{3pt}\noindent
{\sl {\bfseries Analysis of Adjusted Graphs.\hspace{3pt}}}
Table \ref{tbl:outdegree} shows the statistics of the graphs for all datasets. The mean top 5\% outdegree represents the mean outdegree for the top 5\% nodes ranked in descending order of outdegree. The mean bottom 5\% indegree represents the mean indegree for the bottom 5\% of nodes ranked in descending order of indegree. Therefore, it is expected that the two represent the trends of higher outdegrees and lower indegrees, respectively. The two were roughly adjusted by the specified expected degree parameters in Table \ref{tbl:optimized-edge}, especially for SA and SAC. The mean indegree distance represents the mean length of edges in graphs, which are transposed and pruned to only the 10 shortest edges only to compute this metric. This metric represents the accuracy of the incoming edges for each node of the optimized graphs. For SAC, the mean indegree distances were always longer than the other adjustment types because adding incoming edges under the constraints suppressing outdegrees tends to add longer edges to nodes instead. Thus, it is assumed that the longer edges reduce the precision of SAC. In the following sections, we discuss an analysis on the effectiveness of our individual methods in detail.

\subsection{Effectiveness of Each Method}
\noindent{\sl 
{\bfseries Tree-based Index.\hspace{3pt}}}
Our methods use the tree-based index of the NGT to find the near neighboring nodes close to a query as seed nodes to explore a graph. To distinguish the effectiveness of the tree-based index from those of our proposed methods, we clarified the effectiveness of the tree-based index. Each leaf node of the tree has up to 100 objects, which are the nodes of a graph. The 10 nearest objects neighboring the vantage point of a leaf node were used as the seed nodes. To compare with the case of not using the tree-based index, we evaluated the search process with the 10 seed nodes, which were randomly chosen from all of the nodes in the graph instead of using the tree-based index. Fig.~\ref{fig:path-adjust-effect}(a) shows the number of distance computations versus precision when using random and tree-based seeds with DA for all datasets. For lower dimensional datasets, SIFTs and GLOVE 1M, it was assumed that, since exploring the graphs is efficient enough, large improvements due to the tree-based index did not appear. The number of distance computations of GLOVE 2M was significantly improved compared with that of GIST. It is assumed that since the dimension of GLOVE 2M was lower than that of GIST and that the distortion of GLOVE 2M did not come from the dimensionality, nodes closer to a query object could be effectively found by using the tree-based index. However, the improvement of GIST was very slight because the neighboring seed nodes with the tree-based index did not improve the exploring of a graph due to its higher dimensionality. Table \ref{tbl:tree-effect} shows the number of distance computations and precision of the seed nodes when using the tree-based index. The number of distance computations was very small and depended on the number of all objects, i.e., the depth of the tree structure. Even though precision was extremely low, the search results with the tree-based index were effective as the seed nodes to explore a graph. 

\begin{table}
\begin{center}
\caption{Number of distance computations and precision with tree-based index}
\label{tbl:tree-effect}
\vspace{-3mm}
\scriptsize
\begin{tabular}{|l|c|c|c|c|c|}
\hline
Dataset & SIFT & GIST & GLOVE & GLOVE & SIFT \\
      & 1M  &     & 1M     & 2M     & 10M \\
\hline
\hline
\# of computations& 6.7 & 6.3 & 6.3 & 6.8 & 7.8 \\
\hline
Precision & 0.019 & 0.006 & 0.003 & 0.003 & 0.007 \\
\hline
\end{tabular}
\end{center}
\end{table}

\vspace{3pt}\noindent
{\sl {\bfseries Path Adjustment.\hspace{3pt}}}
Fig.~\ref{fig:path-adjust-effect}(b) shows improvements with our path adjustment with DA for all datasets. The term ``npa'' in Fig.~\ref{fig:path-adjust-effect}(b) and \ref{fig:dgree-histogram} represents cases in which the path adjustment was not applied. There was more or less an improvement for all datasets. This trend can also be seen for SA and SAC. Moreover, the path adjustment reduced not only the number of computations but also the outdegrees significantly. Fig.~\ref{fig:path-adjust-effect}(c) shows the reduction in the outdegrees with the path adjustment for DA. The path adjustment almost halved the outdegrees. This reduction also occurred for SA and SAC. Since the reduction in outdegrees can reduce the memory usage for edges, path adjustment is effective in reducing the memory usage for indexing. Fig.~\ref{fig:dgree-histogram} shows an (a) outdegree and (b) indegree histogram for GIST. Since the curves shifted to the left due to the path adjustment, the path adjustment reduced both the indegrees and outdegrees for all nodes. Therefore, the reduction in outdegrees reduces the number of computations. However, since the path adjustment makes some nodes have indegrees of less than 10 for SA and SAC, precision might be reduced. Since there are clearly fewer such nodes for DA than for SA and SAC, the reduction in precision can be suppressed. Moreover, since the outdegrees of SAC without path adjustment concentrate on only a few outdegree values around the specified expected outdegree $e_o$, the constraints of SAC are effective.

\vspace{3pt}\noindent
{\sl {\bfseries Dynamic Degree Adjustment.\hspace{3pt}}}
Fig.~\ref{fig:dgree-histogram}(c) shows the effectiveness of our dynamic degree adjustment for GIST and the number of distance computations versus precision for varying dynamic degree-adjustment parameters with optimized $e_o=10$ and $e_i=115$. Parameters $e_0=30$ and $w_e=20$ resulted in the smallest number of distance computations. Since this tendency was almost the same as those of the other datasets, we used these values as the parameters in our experiments. The figure also shows that DA without our dynamic degree adjustment significantly increased the number of distance computations compared with DA for the same graph. Therefore, our dynamic degree adjustment can effectively reduce the number of distance computations.

\begin{figure*}
\begin{center}
\subfigure[SIFT 1M] 
{\includegraphics[height=3.2cm] {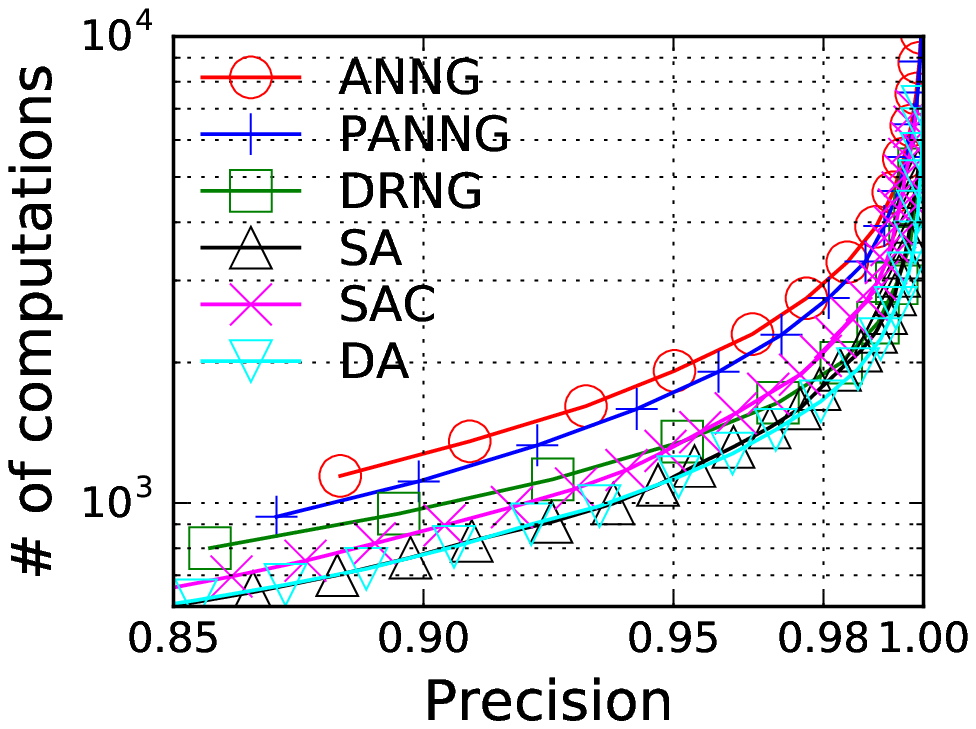}}
\hspace{0mm}
\subfigure[GIST] 
{\includegraphics[height=3.2cm] {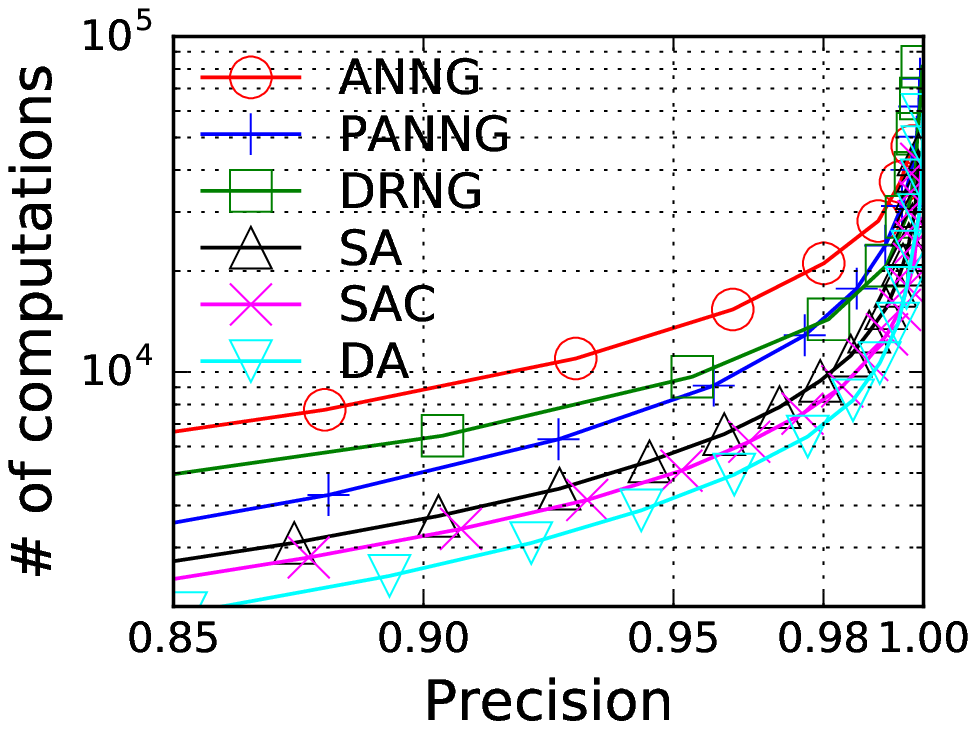}}
\hspace{0mm}
\subfigure[GLOVE 1M] 
{\includegraphics[height=3.2cm] {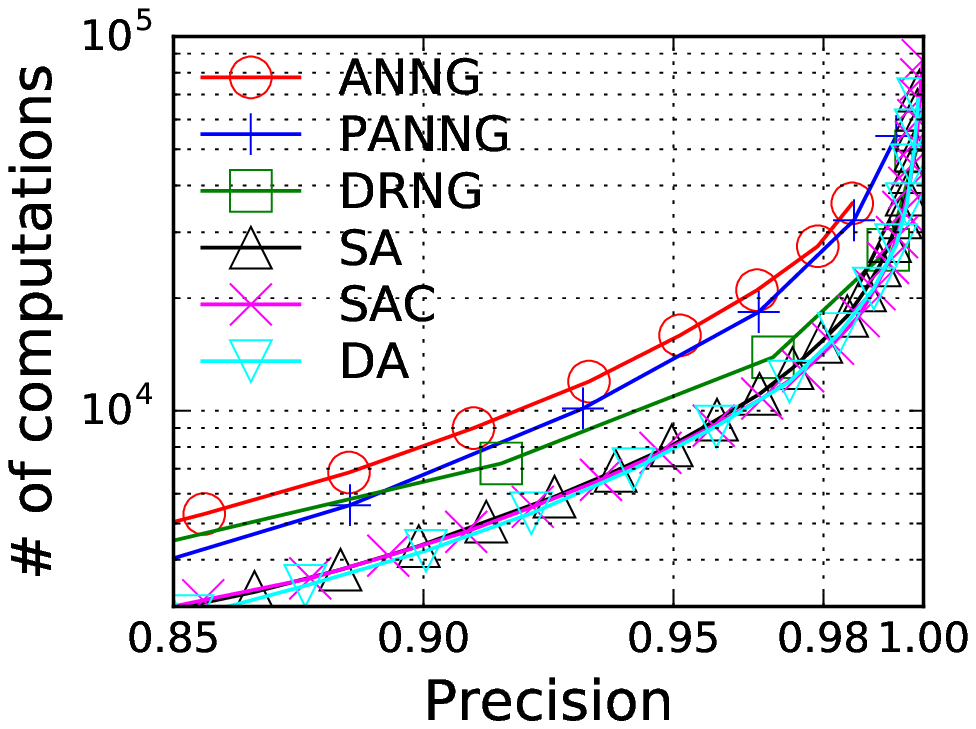}}
\hspace{0mm}
\subfigure[GLOVE 2M] 
{\includegraphics[height=3.2cm] {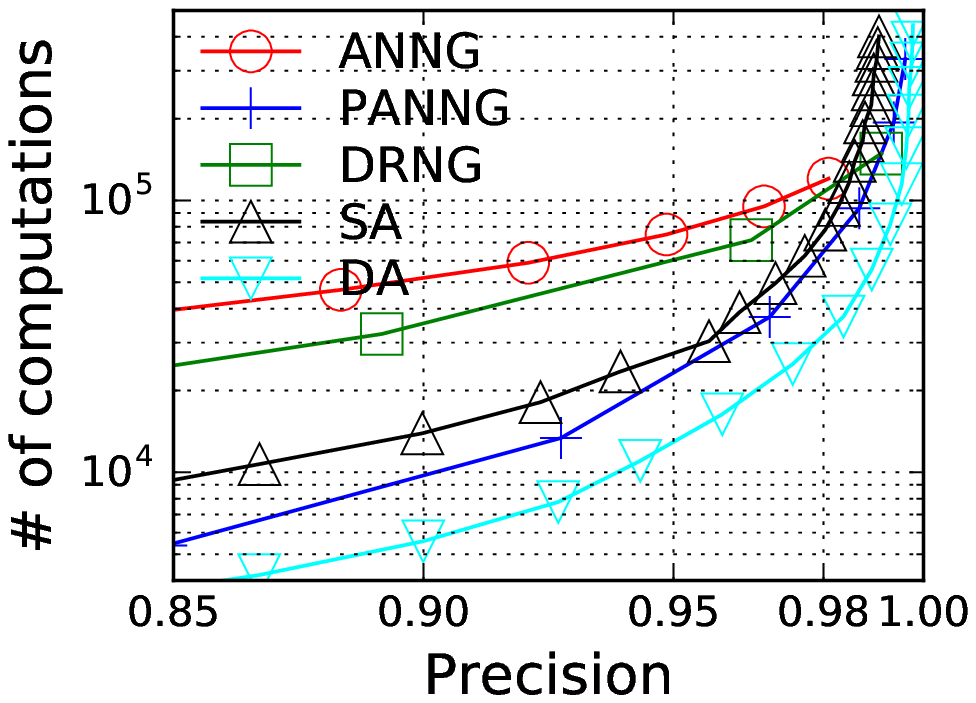}}
\caption{Number of distance computations vs. precision. Comparison of our methods with KNNG-based indexes.}
\label{fig:dist-precision-knng}
\vspace{0mm}
\end{center}
\end{figure*}

\begin{figure*}
\begin{center}
\subfigure[SIFT 1M] 
{\includegraphics[height=3.2cm] {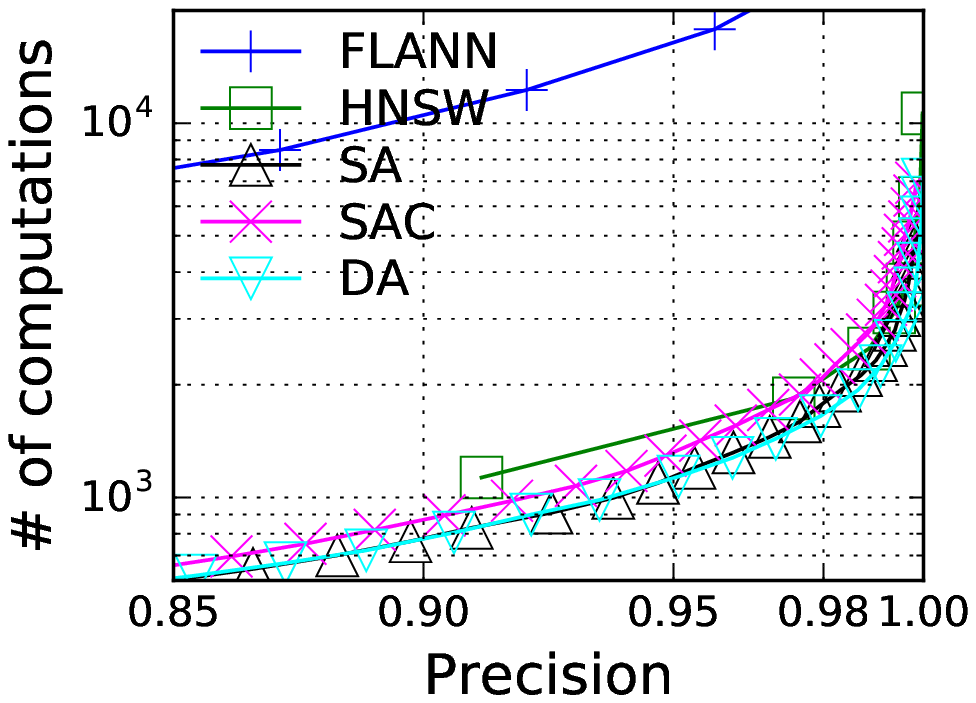}}
\hspace{0mm}
\subfigure[GIST] 
{\includegraphics[height=3.2cm] {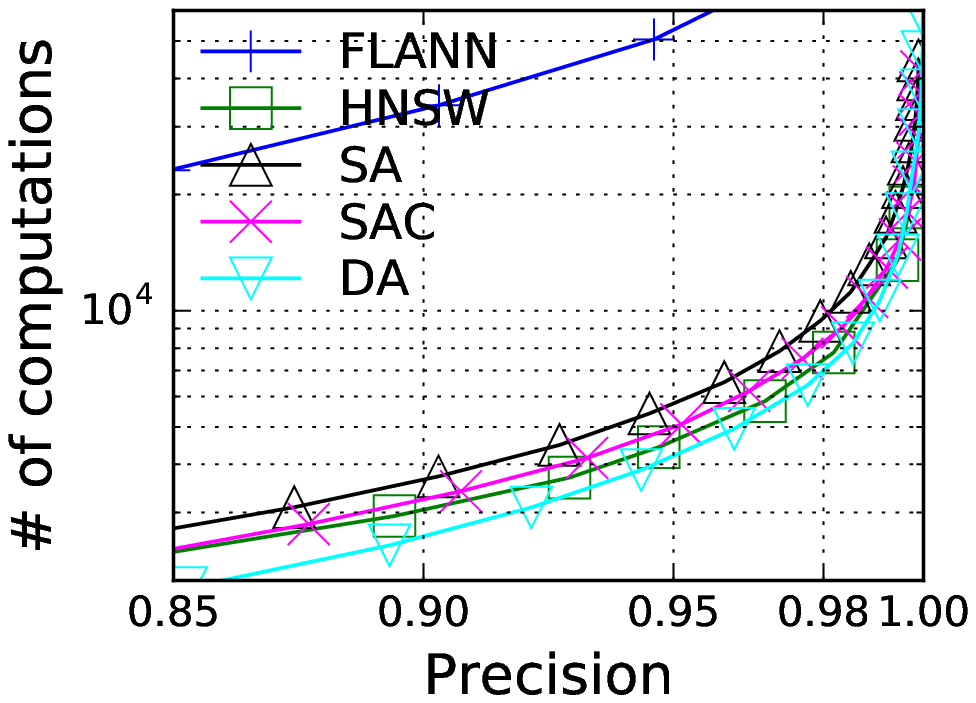}}
\hspace{0mm}
\subfigure[GLOVE 1M] 
{\includegraphics[height=3.2cm] {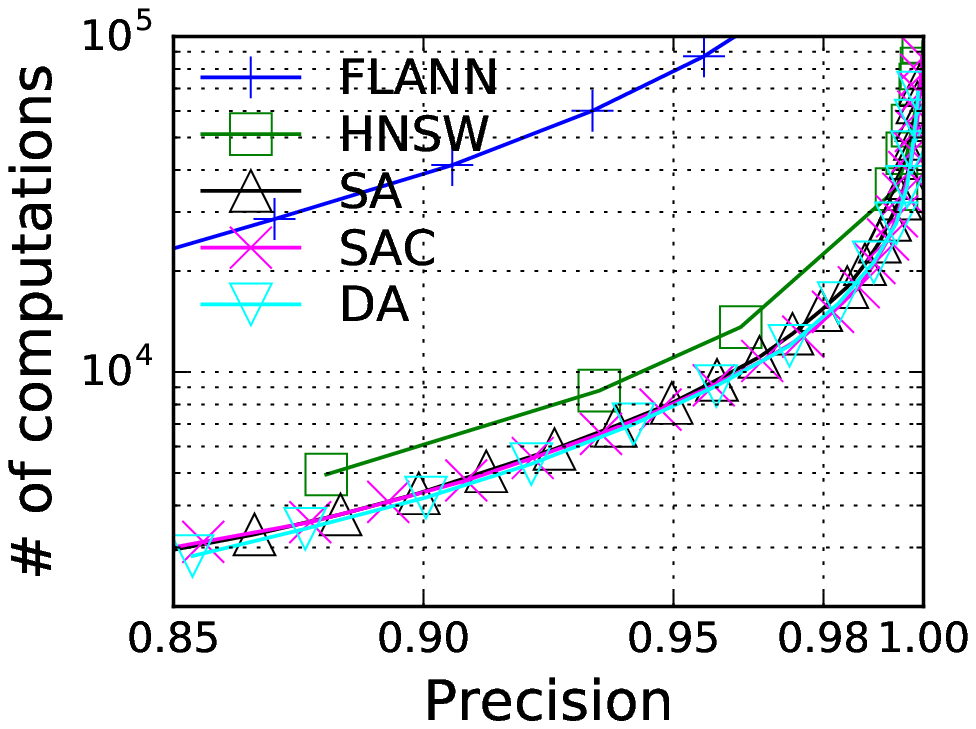}}
\hspace{0mm}
\subfigure[GLOVE 2M] 
{\includegraphics[height=3.2cm] {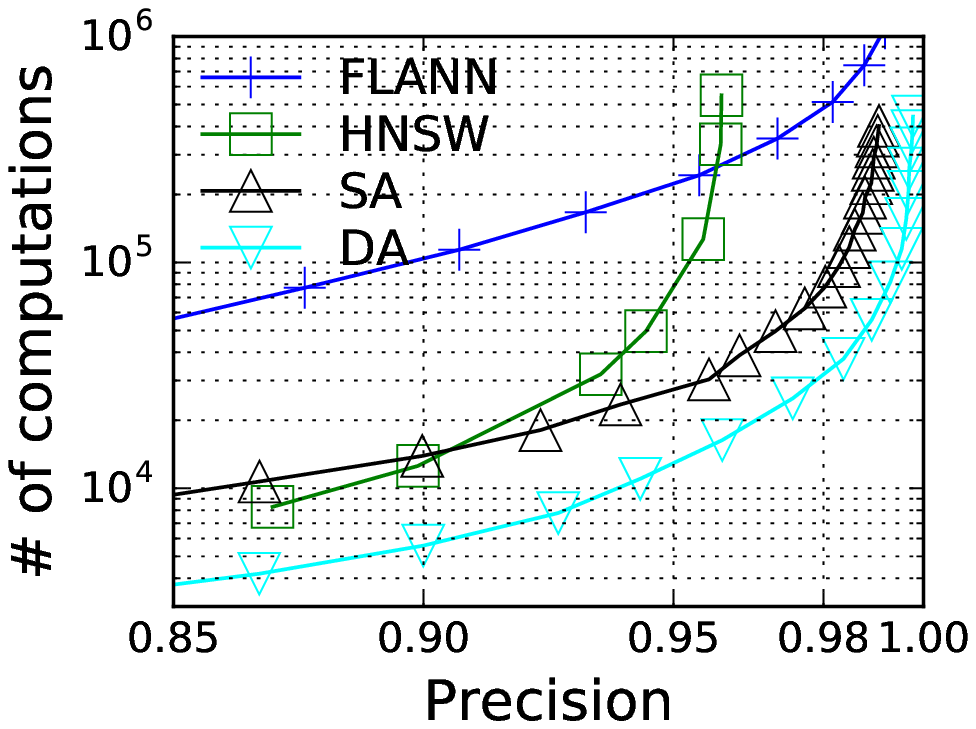}}
\caption{Number of distance computations vs. precision. Comparison with non-KNNG-based indexes.}
\label{fig:dist-precision-non}
\vspace{0mm}
\end{center}
\end{figure*}

\begin{figure*}
\begin{center}
\subfigure[SIFT 1M] 
{\includegraphics[height=3.2cm] {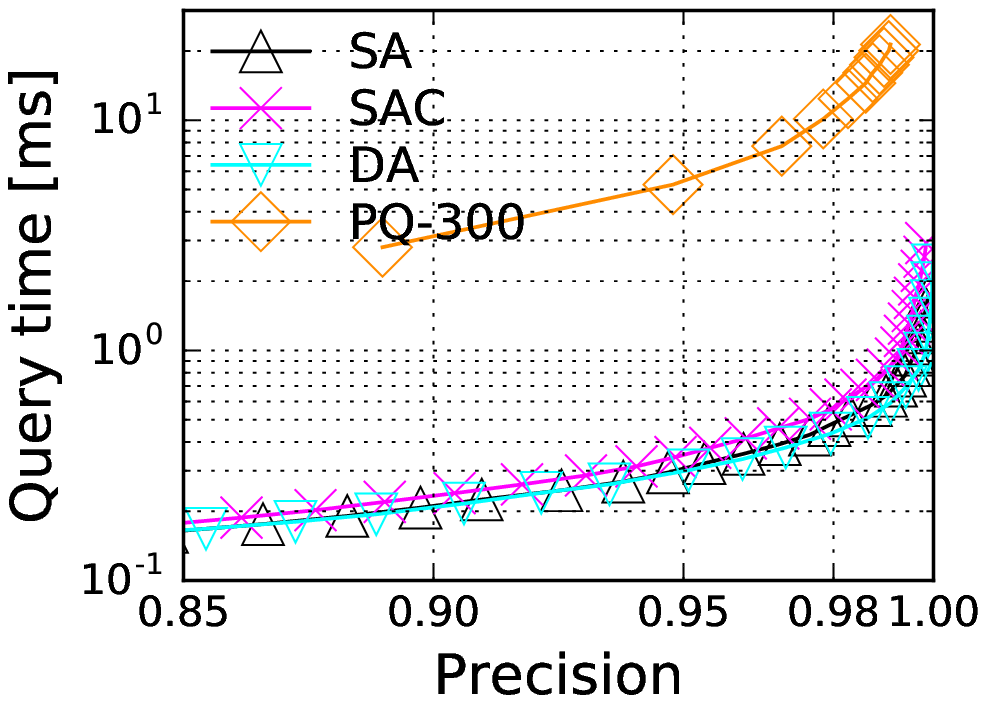}}
\hspace{0mm}
\subfigure[GIST] 
{\includegraphics[height=3.2cm] {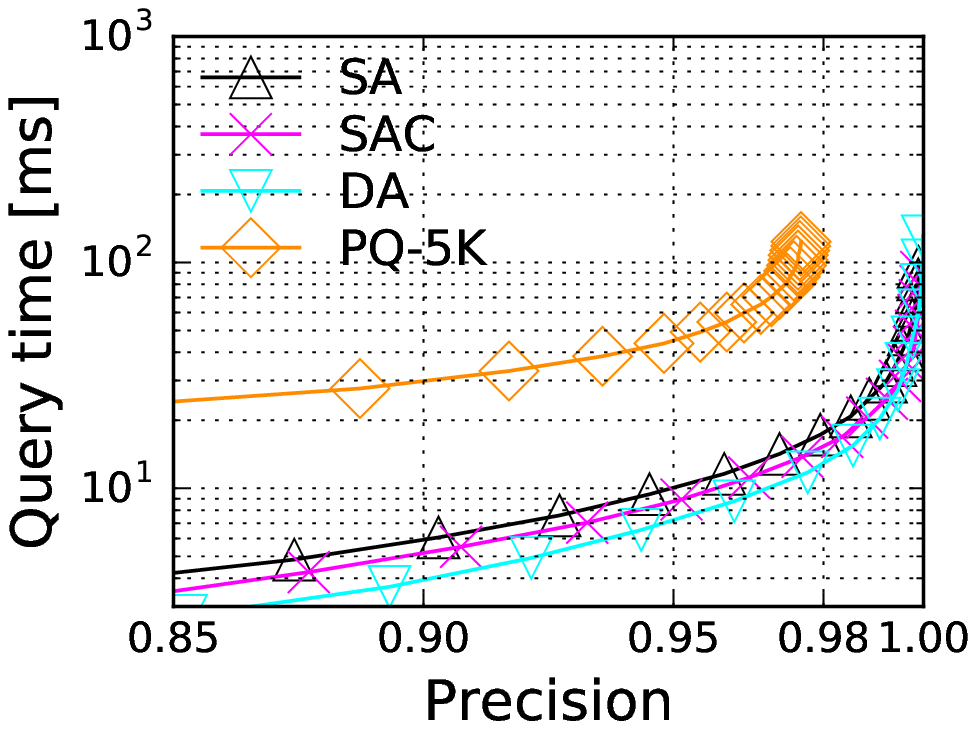}}
\hspace{0mm}
\subfigure[GLOVE 1M] 
{\includegraphics[height=3.2cm] {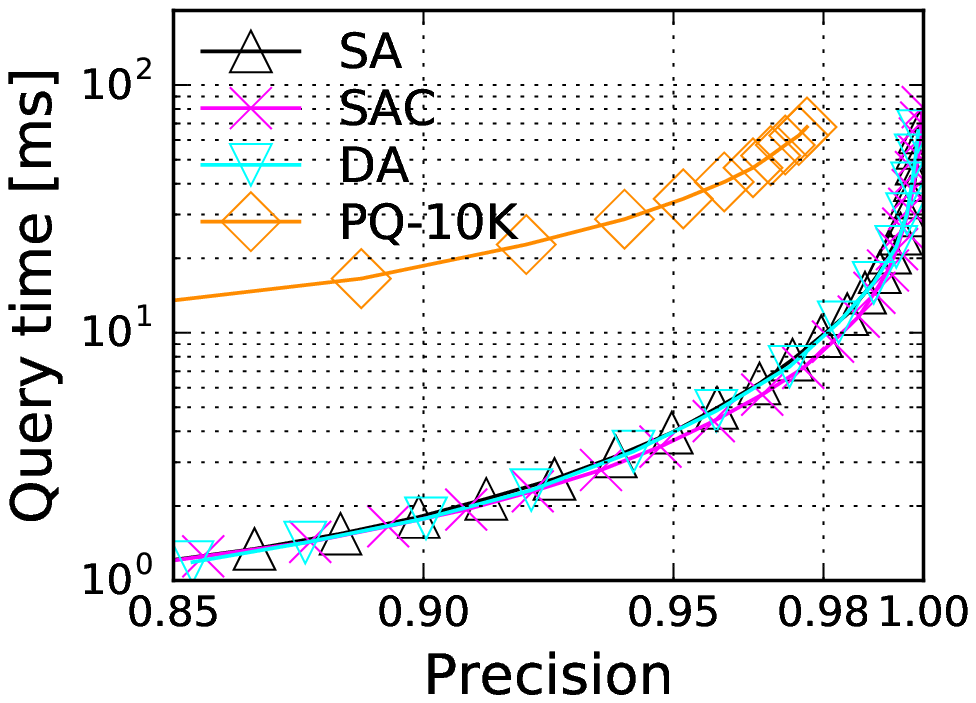}}
\hspace{0mm}
\subfigure[GLOVE 2M] 
{\includegraphics[height=3.2cm] {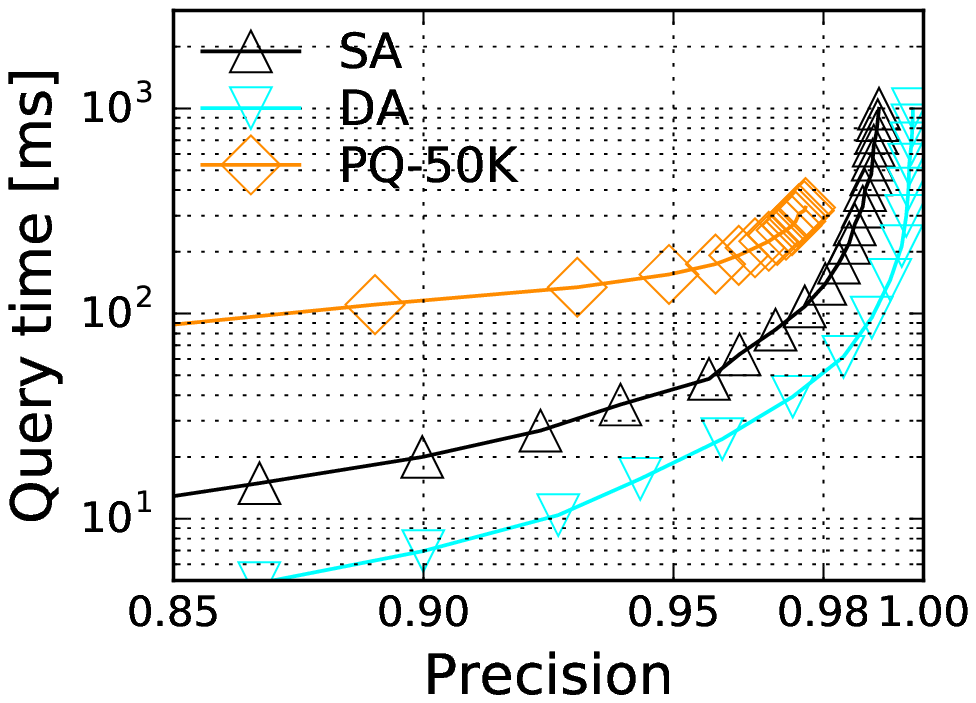}}
\caption{Query time vs. precision. Comparison with PQs. Suffixes of PQs represent $k^{\prime\prime}$ for verification step.}
\label{fig:time-precision-non}
\vspace{0mm}
\end{center}
\end{figure*}

\begin{table}
\begin{center}
\caption{HNSW parameters}
\label{tbl:hnsw-param}
\vspace{-3mm}
\scriptsize
\begin{tabular}{|l|c|c|c|c|c|}
\hline
Parameter & SIFT & GIST & GLOVE & GLOVE & SIFT \\
           & 1M  &      & 1M     & 2M     & 10M \\  
\hline
{\scriptsize $efConstruction$} & 400 & 400 & 800 & 800 & 400 \\
\hline
$M$ & 32 & 32 & 48 & 48 & 32 \\
\hline
\end{tabular}
\vspace{0mm}
\end{center}
\end{table}

\subsection{Comparison with Existing Methods}
\noindent{\sl {\bfseries Comparison among KNNG-based Indexing Methods.\hspace{3pt}}}
Fig.~\ref{fig:dist-precision-knng} shows comparisons with other KNNG-based indexing methods for the 1M and 2M datasets. To compare in terms of graph structures, all indexes used Algorithm \ref{alg:knnsearch} with the tree-based index for fair comparison. The parameters of the ANNG and PANNG were set to the best values ($k_c=10, k_r=30$ and $k_p=60$), which an experiment with the PANNG \cite{iwasaki2016pruned} showed. The DRNG was constructed from the AKNNG with an outdegree of 200. Our DA outperformed the others throughout the entire target precision range for all datasets. Around a precision of 1.0 for SIFT 1M and GLOVE 2M, the numbers of computations for SA and SAC were more than that for DRNG, indicating that DRNG is effective for high precision. It is also assumed that this is because our target precision range was not around 1.0 but [0.9, 0.98].

\vspace{3pt}\noindent{\sl {\bfseries Comparison with Non-KNNG-based Indexing Methods.\hspace{3pt}}}
Fig.~\ref{fig:dist-precision-non} shows comparisons with FLANN\footnote{https://www.cs.ubc.ca/research/flann, v1.8.4} and HNSW\footnote{https://github.com/searchivarius/nmslib, v1.7.3.4} as non-KNNG-based indexing methods. Although HNSW is a graph-based indexing method since it is a hierarchical index unlike our graphs, our KNN search could not be applied to HNSW. FLANN automatically selected the best algorithm for the dataset and target precision we specified, i.e., 0.95. It selected hierarchical k-means partitioning for all datasets. For HNSW, we readjusted its parameters in consideration of its benchmark, as Table \ref{tbl:hnsw-param} shows. HNSW's curves were plotted by varying the parameter $efSearch$. Our DA outperformed FLANN and HNSW. For SIFT 1M, GIST, and GLOVE 1M, HNSW was close to DA. However, the precision of HNSW for GLOVE 2M could not even reach 0.97 since HNSW was not effective for GLOVEs, and, in addition, the dimensionality of GLOVE 2M was three times higher than that of GLOVE 1M.

Fig.~\ref{fig:time-precision-non} shows comparisons with the product quantization-based method (PQ) \cite{jegou2011product} in terms of query time since the PQ could not be compared in terms of distance computations because it does not compute distances with original objects. While it does not require objects in memory, the search accuracy is significantly lower. To obtain our target precision, we added a verification step after the PQ search, which computes distances for the results of the PQ by using the objects in memory and returns the $k$ nearest neighbors. Let the number of objects that are passed from the PQ to the verification step be $k^{\prime\prime}$, where the precision increases as $k^{\prime\prime}$ increases. From a preliminary experiment on the PQ, parameters that were almost the best were determined. We used the number of code words $k'=1024$ for the coarse quantizer, numbers of subvectors $m=16$ for SIFTs and GIST, $m=10$ for GLOVE 1M, $m=15$ for GLOVE 2M, and the number of code words for the product quantizer $k^{*}=256$, which can produce almost the shortest query time. The curves of the PQ were plotted by varying the number of the nearest neighbors of the coarse quantizer $w$. From the figures, the PQ's query times are clearly longer than our proposed methods for the high precision range.

\begin{figure*}
\begin{center}
\subfigure[] 
{\includegraphics[height=3.2cm] {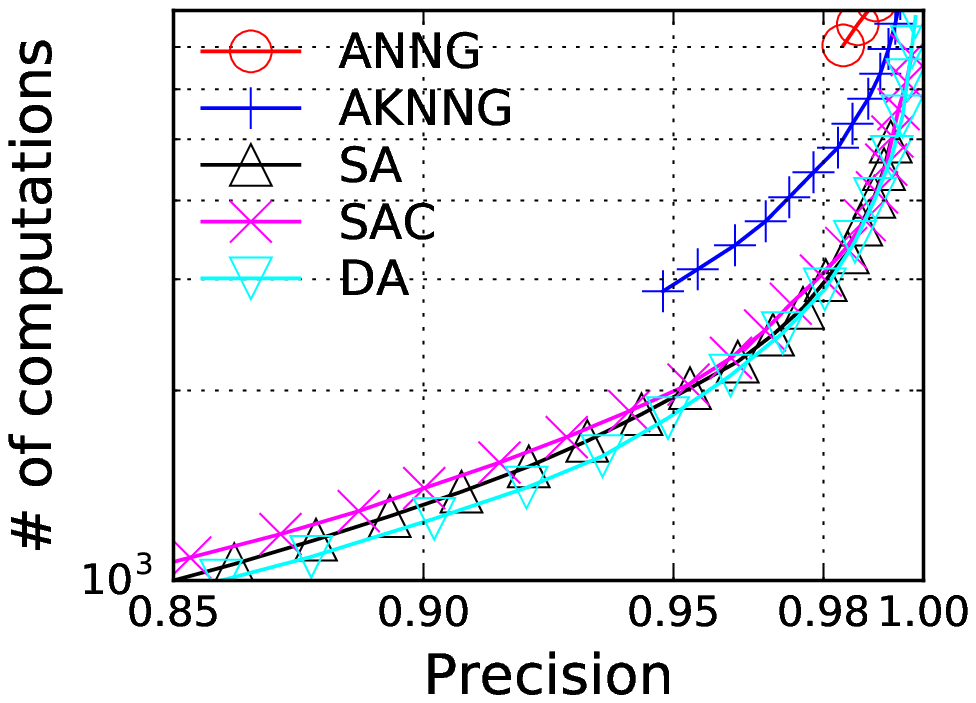}}
\hspace{0mm}
\subfigure[] 
{\includegraphics[height=3.2cm] {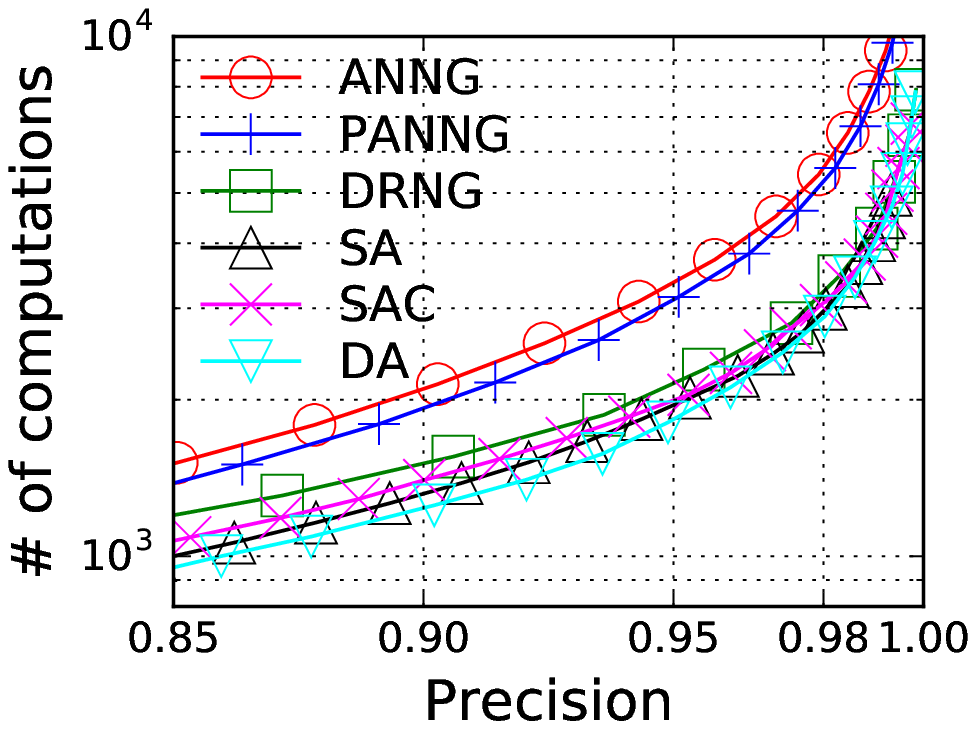}}
\hspace{0mm}
\subfigure[] 
{\includegraphics[height=3.2cm] {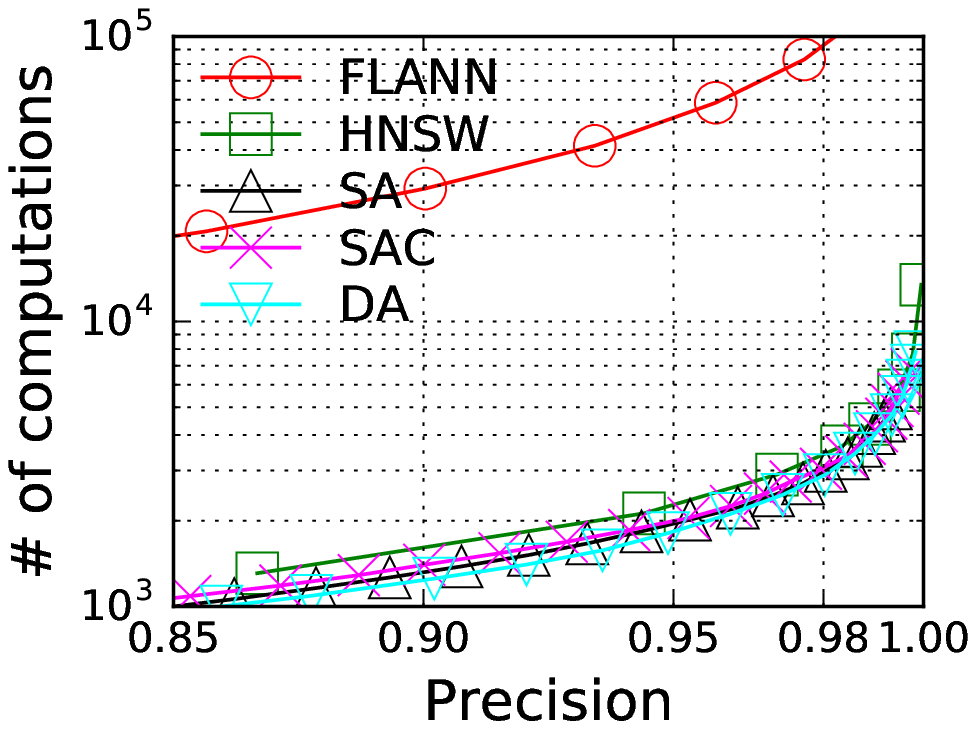}}
\hspace{0mm}
\subfigure[]
{\includegraphics[height=3.2cm] {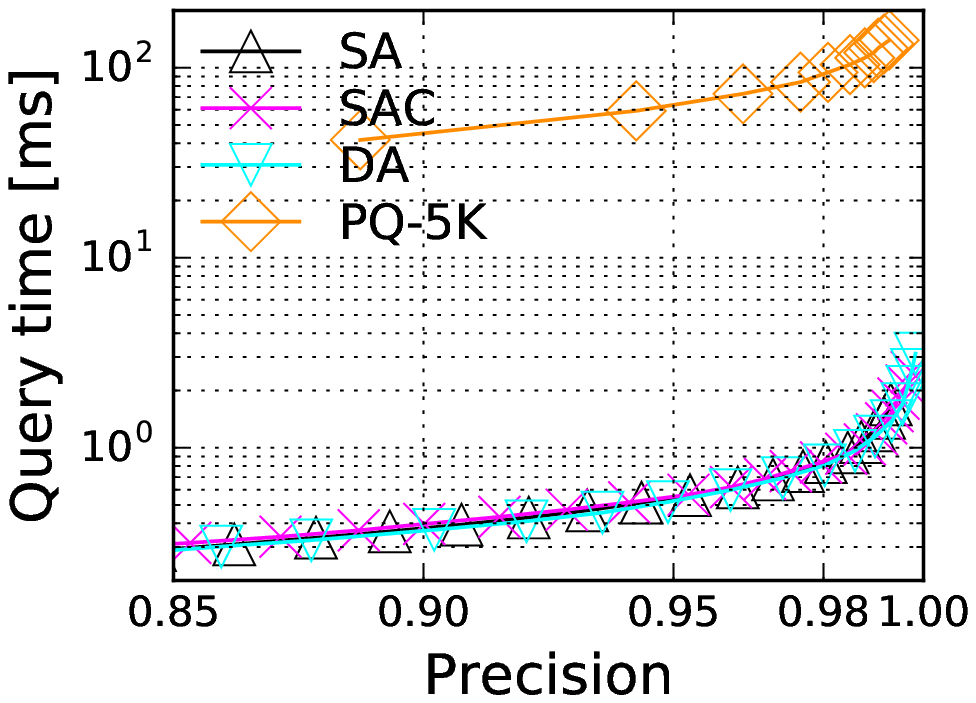}}
\caption{Number of distance computations vs. precision for SIFT 10M. Comparison (a) among our methods, (b) with KNNG-based indexes, and (c) with non-KNNG-based indexes. (d) Query time vs. precision. Comparison with PQs.}
\label{fig:time-precision-3}
\vspace{0mm}
\end{center}
\end{figure*}

\begin{figure*}
\begin{center}
\subfigure[] {
\includegraphics[height=3.2cm] {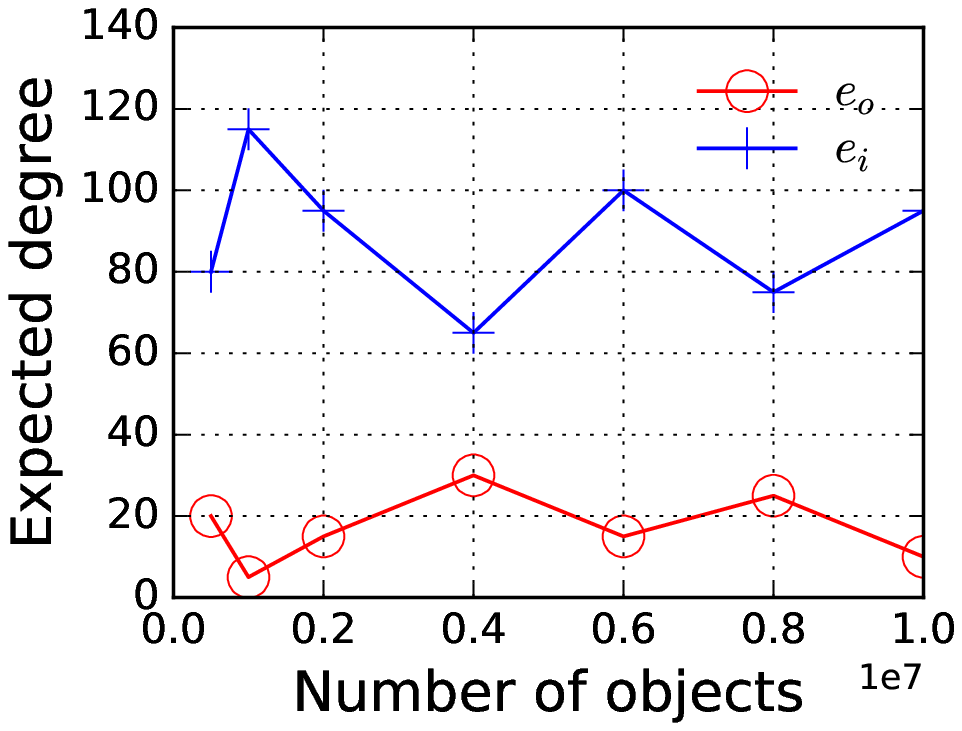}
}
\hspace{5mm}
\subfigure[] {
\includegraphics[height=3.2cm] {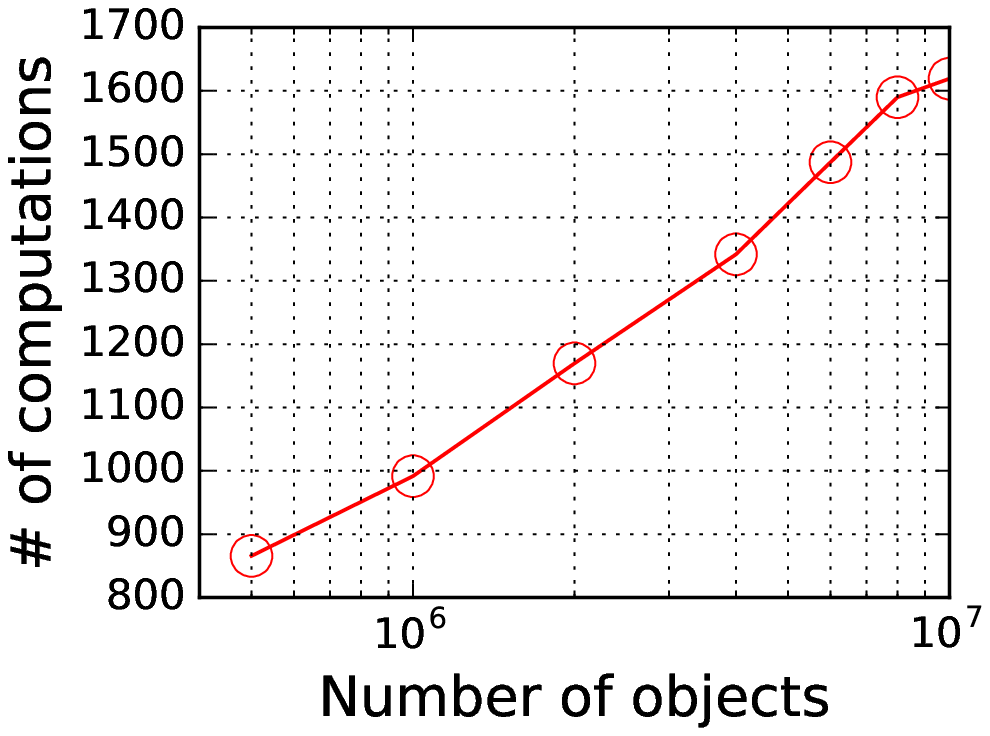}
}
\hspace{5mm}
\subfigure[] {
\includegraphics[height=3.2cm] {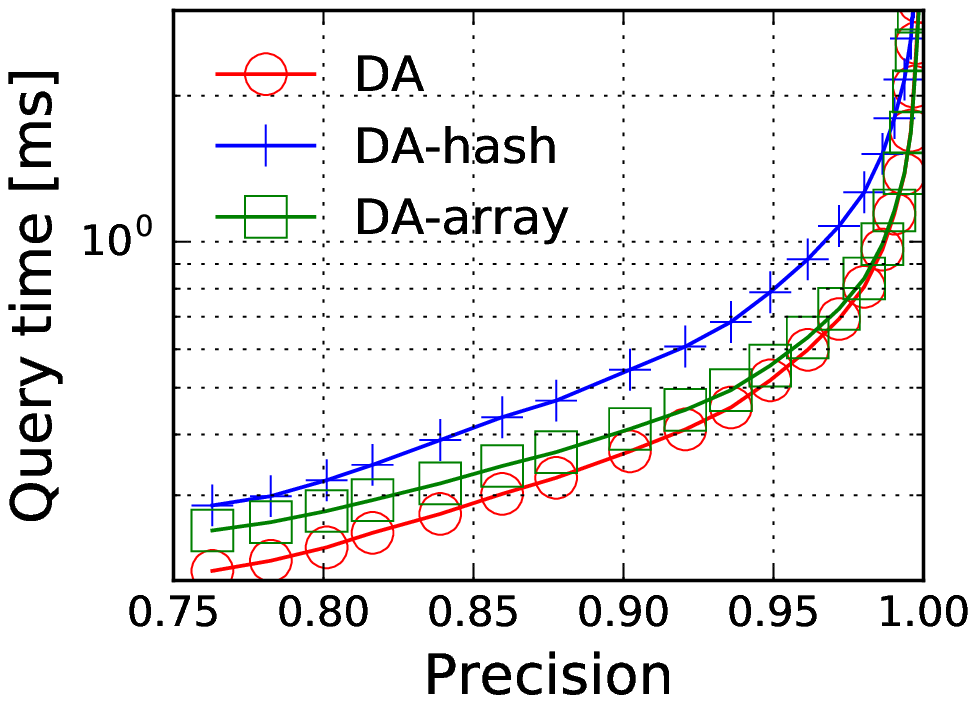}
}
\caption{(a) Optimal expected indegree and outdegree parameters vs. number of indexed SIFT objects. (b) Number of distance computations vs. number of indexed SIFT objects. (c) Query time vs. precision with DA with three types of visited-node management for SIFT 10M.}
\label{fig:number-of-object-effect}
\vspace{0mm}
\end{center}
\end{figure*}

\begin{table*}
\begin{center}
\caption{Construction times and memory usage}
\label{tbl:construction}
\vspace{-3mm} 
\begin{tabular}{|c|c|c|c|c|}
\hline
ANNG and & Static degree & Static degree & Path & Memory \\
AKNNG& adjustment & adjustment & adjustment & usage \\
construction & & with constraint & & \\
\hline\hline
250.9 min & 5.6 min & 13.8 min & 158.0 min & 9.8 GB \\
\hline
\end{tabular}
\vspace{0mm}
\end{center}
\end{table*}

\subsection{Experimental Results for Large Dataset}
Fig.~\ref{fig:time-precision-3} shows comparisons among our methods, with KNNG-based indexing methods, and with non-KNNG-based indexing methods in terms of the number of computations and with non-KNNG-based indexing methods in terms of the query time for SIFT 10M in the same manner for the 1M and 2M datasets discussed above. The trend was almost the same as that for SIFT 1M, that is to say, DA outperformed SA, and SAC and outperformed the previous indexing methods for a large dataset in terms of the number of computations and query time. 

Fig.~\ref{fig:number-of-object-effect}(a) shows the optimal $e_o$ and $e_i$ versus the number of indexed objects for SIFT 10M with DA. $e_i$ should be higher than $e_o$ for DA. It seemed unstable for the number of indexed objects, unlike $e_o$. However, as Fig.~\ref{fig:degree-optimization}(b) shows, the difference among their numbers did not cause a significant difference in the number of computations. Therefore, the numbers of both edges were considered stable in terms of precision, even though the number of indexed objects increased. Therefore, to obtain the best parameters, optimization should be conducted for a graph indexing all target objects. However, from this observation, it is assumed that optimization for a subset of objects can produce almost the best parameters. 

Fig.~\ref{fig:number-of-object-effect}(b) shows the number of distance computations versus that of indexed objects with the optimal parameters shown in Fig.~\ref{fig:number-of-object-effect}(a) for each number of indexed objects. The computational complexity of the search was $O(\log n)$ due to the almost straight line on a logarithmic scale. 

\vspace{3pt}\noindent{\sl {\bfseries Effectiveness of Improved Visited-Node Management.\hspace{3pt}}}
Fig.~\ref{fig:number-of-object-effect}(c) shows the effectiveness of search acceleration with our improved visited-node management. In the figure, DA-hash used an unordered map container of the C++ Standard Library, DA-array used a simple array, and DA used our improved visited-node management. The management outperformed both the unordered map container and simple array. However, the reduction in query time from using the management was not large for the high precision range compared with using a simple array. Most of the time of using an array is occupied by a fixed initialization time of zero for the array because checking visited nodes takes an extremely short amount of time. Therefore, the rate of the initialization time for the entire query time for the high precision range was smaller than that for the low precision range. From this observation, it is expected that if more than 10 M objects are stored to a graph, a further reduction in the query time with our visited-node management is possible compared with that with the simple array.

\vspace{3pt}\noindent{\sl {\bfseries Construction Time and Memory Usage.\hspace{3pt}}}
Table \ref{tbl:construction} shows the processing times of AKNNG and ANNG constructions, degree adjustments, and path adjustments for SIFT 10M. The ANNG constructions were conducted in parallel by using the NGT. Since others were not processed in parallel, these processing times can be reduced by modifying the algorithms running in parallel. All of the processing times basically depend on the number of edges that they process. Although the ANNG construction times were long due to generating 200 edges for each node, only an optimized number of edges should be generated for each dataset to reduce the construction times of ANNGs. Table \ref{tbl:construction} also shows the memory usage including that for storing edge lengths for all edges and the management data used by the NGT. In this experiment, all objects were stored as a 4-byte floating point number even for the 1-byte SIFT for fair comparison.

\section{Conclusion}
To improve the query time with a graph, we proposed three degree-adjustment methods for adjusting the indegrees and outdegrees for each node in the graph: static degree adjustment, static degree adjustment with constraints, and dynamic degree adjustment. We also proposed a path adjustment for optimizing a graph in consideration of the search path, and we improved the managing of nodes visited during the search process. We also showed that most of our proposed methods outperformed previous methods for various sorts of datasets. Moreover, our static degree adjustment, dynamic degree adjustment, path adjustment, and visited-node management were each effective for different sorts of datasets, improving the search performance. It is assumed that all of these methods are indispensable for application to various sorts of datasets. We also showed how to automatically optimize the parameters of our degree adjustment to construct optimal graphs. The source code of our proposed methods is included in NGT and is available to the public.

\appendix
\section*{ANNG Construction}
While each object is incrementally added to the ANNG, neighboring nodes to the added node are searched for using the partially constructed ANNG to reduce the construction cost. Algorithm \ref{alg:ConstructANNG} shows the construction algorithm. Let $O, k_c, \epsilon_c$, and $G_a$ be a set of inserted objects, number of edges, $\epsilon$ for KNN search during construction, and resultant ANNG, respectively.

\begin{algorithm}
\begin{footnotesize}
\caption{ConstructANNG}
\label{alg:ConstructANNG}
\begin{algorithmic}[1]
\REQUIRE $O, k_c, \epsilon_c$
\ENSURE $G_a(V_a, E_a)$
\STATE $V_a \leftarrow O$
\FORALL{$o \in O$}
\STATE $N(G_a, o) \leftarrow \mathrm{KnnSearch}(G_a, o, k_c, \epsilon_c, \infty)$
\FORALL{$n \in N(G_a, o)$}
\STATE $N(G_a, n) \leftarrow N(G_a, n) \cup \{ o\} $
\ENDFOR
\ENDFOR
\RETURN $G_a$
\end{algorithmic}
\end{footnotesize}
\end{algorithm}

\ifCLASSOPTIONcaptionsoff
  \newpage
\fi



\bibliographystyle{plain}
\bibliography{index} 

\end{document}